\documentclass[12pt,a4paper]{article}
\usepackage{amsmath}
\usepackage{graphicx}
\usepackage{amsfonts}
\usepackage{amssymb}
\usepackage{epsfig}
\usepackage{psfrag}
\usepackage{wrapfig} 
\usepackage{color}

\definecolor{yellow}{rgb}{1,0.9,0} % color values Red, Green, Blue
\definecolor{wine}{rgb}{0.5,0,0.4}
\definecolor{ancientrose}{rgb}{0.7,0,0.4}
\definecolor{cream}{rgb}{1.,1.,0.7} 
\definecolor{violet}{rgb}{1.,0.9,0.95} 
\definecolor{lightgreen}{rgb}{0.8,1,0.8} 
\def\nue{\mathrel{{\nu_e}}}
\def\numu{\mathrel{{\nu_\mu}}}
\def\nutau{\mathrel{{\nu_\tau}}}
\def\nux{\mathrel{{\nu_x}}}

\def\barnue{\mathrel{{\bar \nu}_e}}  
\def\barnumu{\mathrel{{\bar \nu}_\mu}}  
\def\barnutau{\mathrel{{\bar \nu}_\tau}}

\def \lta {\mathrel{\vcenter{\hbox{$<$}\nointerlineskip\hbox{$\sim$}}}}
\def \gta {\mathrel{\vcenter{\hbox{$>$}\nointerlineskip\hbox{$\sim$}}}}

\begin{document}

%\pagecolor{cream} % Background color
%\color{black}     % Text color

\begin{titlepage}

\vspace*{0.2cm}
\begin{center}

{\Large \bf Supernova neutrinos: difference of $\numu$ -- $\nutau$ fluxes 
and conversion effects} \\

\vspace{0.8cm}
{\large Evgeny Kh. Akhmedov$^{a,1}$, Cecilia Lunardini$^{b,2}$ and
Alexei Yu. Smirnov$^{c,d,3}$ } \\
\vspace{0.3cm}
{\em $^{a}$ Centro de F\'\i sica das Interac\c c\~oes Fundamentais (CFIF)  
Departamento de F\'\i sica, Instituto Superior T\'ecnico  Av. Rovisco Pais, 
P-1049-001 Lisboa, Portugal} \\

\vspace{0.2cm}
{\em $^{b}$ Institute for Advanced Study, Einstein drive, 08540 Princeton, 
New Jersey, USA}

\vspace{0.2cm}
{\em $^{c}$ The Abdus Salam ICTP, Strada Costiera 11, 34100 Trieste, Italy}

\vspace{0.2cm}
{\em $^d$ Institute for Nuclear Research, RAS, Moscow 123182, Russia.}

%\maketitle \vspace{2cm}
\end{center} 
\begin{abstract}
\noindent 
The formalism of flavor conversion of supernova neutrinos is generalized to 
include possible differences in the fluxes of the muon and tau neutrinos 
produced in the star. In this case the radiatively induced difference of 
the  $\nu_{\mu}$ and $\nu_{\tau}$ potentials in matter  becomes important.  
The $\nu_{\mu}$ and $\nu_{\tau}$ flux differences 
can manifest themselves in the  
effects of the Earth matter on the observed $\nue$ ($\barnue$) signal if: 
(i) the neutrino mass hierarchy is normal (inverted); (ii) the solution of 
the solar neutrino problem is in the LMA region; (iii) the mixing $U_{e3}$ 
is relatively large: $|U_{e3}|\gta 10^{-3}$. We find that for differences
in the $\numu$ -- $\nutau$ ($\barnumu$ -- $\barnutau$) average energies
and/or integrated luminosities $\lta 20\%$, the relative deviation of the 
observed $\nue$ ($\barnue$) energy spectrum at  $E \gta 50$ MeV  from that in  
the case of the equal fluxes can reach $\sim 20 - 30 \%$ ($\sim 10 - 15
\%$) for neutrinos crossing the Earth. It could be detected in future if
large detectors sensitive to the $\nue$ ($\barnue$) energy spectrum become
available. The study of this effect would allow one to test the predictions 
of the $\numu$, $\nutau$, $\barnumu$, $\barnutau$ fluxes from supernova
models and therefore give an important insight into the properties of matter 
at extreme conditions. It should be taken into account in the reconstruction 
of the neutrino mass spectrum and mixing matrix from the supernova neutrino 
observations. We show that even for unequal $\numu$ 
and $\nutau$ fluxes, effects of leptonic CP violation can not be studied in
the supernova neutrino experiments. 

\end{abstract}

\vskip10pt
\noindent
{\it PACS:} 14.60.Pq, 97.60.Bw.

\noindent {\it Keywords:} neutrino conversion; matter effects; supernova.
\vfil
\noindent
\footnoterule  {\small $^1$On leave from
National Research Center Kurchatov Institute, Moscow 123182, Russia.
\vskip-1pt\noindent}{\small
E-mail: akhmedov@cfif.ist.utl.pt\vskip-1pt\noindent} {\small
$^{2}$E-mail: lunardi@ias.edu\vskip-1pt\noindent} 
{\small $^3$ E-mail: smirnov@ictp.trieste.it\vskip-1pt\noindent}

\thispagestyle{empty}
\end{titlepage}

\setcounter{page}{1}

%%%%%%%%%%%%%%%%%%%%%%%%%%%%%%%%%%%%%%%%%%%%%%%%%%%%%%%%%%%%%%%%%%%%%%%%%%
%%%%%%%%%%%%%%%%%%%%%%%%%%%%%%%%%%%%%%%%%%%%%%%%%%%%%%%%%%%%%%%%%%%%%%%%%%%
\section{Introduction}
\label{sect:1}
%%%%%%%%%%%%%%%%%%%%%%%%%%%%%%%%%%%%%%%%%%%%%%%%%%%%%%%%%%%%%%%%%%%%%%%%%%%
The study of supernova neutrinos is of great interest to both
astrophysics and particle physics. As far as the astrophysics is
concerned, the neutrinos provide precious information about the core
gravitational collapse, the mechanism of supernova explosion and the
properties of matter at extreme conditions of large densities and high
temperatures. From the particle physics point of view, supernova
neutrinos allow one to probe the neutrino mixings and mass spectrum.
Indeed, the properties of the neutrino fluxes are modified by flavor
conversion  effects inside the star (see 
\cite{Mikheev:1986if}-\cite{Minakata:2001cd} as an  incomplete list of 
relevant works)  and in the
matter of the Earth (see \cite{talkalexei} for an  early suggestion and 
\cite{Lunardini:2000sw,Lunardini:2001pb,Takahashi:2001dc} for recent studies).

Until now, the effects of neutrino oscillations and conversion have been 
considered in the assumption that the  fluxes of the muon and tau neutrinos 
produced in the star are identical. Indeed, while the diffusion of $\nue$ in 
the star differs from that of $\numu$ due to the effect of the charged
current  processes, the non-electron flavors,  $\numu$ and $\nutau$, 
and their antineutrinos are expected to undergo similar interaction effects, 
dominated by neutral current scatterings. Therefore, the non-electron 
neutrinos have similar fluxes: 
\begin{equation}
F_\mu^0 \approx  F_{\bar \mu}^0 \approx F_\tau^0 \approx F_{\bar \tau }^0~. 
\label{hier2}
\end{equation} 
In contrast to this, the fluxes of the $\nue$, $\barnue$ and $\numu$ are 
expected to be rather different, with the hierarchy of the 
average energies $\langle E_e \rangle < \langle  E_{\bar e} \rangle < \langle 
E_\mu \rangle~$. 
In calculations of the neutrino transport in the star $\numu$, $\nutau$, 
$\barnumu$ and $\barnutau$ are considered equivalent and treated effectively 
as a single species, $\nu_x$.

Several effects can break the $\numu$ -- $\nutau$ and the $\numu$ -- 
$\barnumu$ (or $\nutau$ -- $\barnutau$) equivalence, giving rise to small
differences between the fluxes of these species. Once effects of weak 
magnetism are considered, the $\numu$ and $\barnumu$ interaction cross section 
in the matter of the star are different; this can lead to a difference between 
the average energies and luminosities of the $\numu$ and $\barnumu$ fluxes as 
large as $\sim 10  \%$ \cite{Horowitz:2001xf}. The production of muons in 
the core of the neutron star results in differences in the transport
of $\numu$ and $\nutau$ inside the star, whose effect on the neutrino fluxes 
exiting the star has not  been fully explored yet \cite{jankapriv}. 
\footnote{Another potential source of  differences between the
$\numu$ and $\nutau$ fluxes are possible effects of new physics, such
as lepton number violating processes or violations of lepton
universality. Such effects could, in principle, influence the whole
dynamics of neutrino transport in supernovae, and their detailed study
goes beyond the scope of the present paper.}

The possibility of studying these phenomena through the observation of a
difference in the $\numu$ and $\nutau$ fluxes makes this difference a quantity 
of particular interest. In addition, if an appreciable $\numu$ -- $\nutau$
difference exists, its effects on the neutrino conversion should be taken into
account while analyzing the supernova neutrino signals in order to 
determine  the neutrino oscillation parameters. 

In this paper we present a generalization of the formalism of neutrino 
conversion in the star and in the Earth to the case of unequal $\numu$ and 
$\nutau$ fluxes. We address the question of whether the $\numu$ --
$\nutau$ flux  differences could be probed via flavor conversion effects
and what the effect of these differences would be on the study of the neutrino 
mixing and mass spectrum with supernova neutrinos.  

The paper is organized as follows. 
After giving some generalities in sec. \ref{sect:2}, the effects of $\numu$ 
-- $\nutau$ flux differences on neutrino conversion in the star are discussed
in sec. \ref{sect:3}. In sec. \ref{sect:4} we study the Earth matter effects 
and comment on possible signatures of the $\numu$ --  $\nutau$ 
flux differences. 
Discussion and conclusions follow in sec. \ref{sect:5}.

%%%%%%%%%%%%%%%%%%%%%%%%%%%%%%%%%%%%%%%%%%%%%%%%%%%%%%%%%%%%%%%%%%%%%%%%%%%
%%%%%%%%%%%%%%%%%%%%%%%%%%%%%%%%%%%%%%%%%%%%%%%%%%%%%%%%%%%%%%%%%%%%%%%%%%%
\section{Supernova neutrinos and different $\numu$ and $\nutau$ fluxes}
\label{sect:2}
%%%%%%%%%%%%%%%%%%%%%%%%%%%%%%%%%%%%%%%%%%%%%%%%%%%%%%%%%%%%%%%%%%%%%%%%%%%

%%%%%%%%%%%%%%%%%%%%%%%%%%%%%%%%%%%%%%%%%%%%%%%%%%%%%%%%%%%%%%%%%%%%%%%%%%%
\subsection{General features}
\label{sect:2.1}

At a given time $t$ from the core collapse the original flux  of the neutrinos
of a given flavor, $\nu_\alpha$, can be described by a ``pinched'' 
Fermi-Dirac spectrum, 
\begin{eqnarray}
F^0_\alpha(E,T_\alpha,\eta_\alpha,L_\alpha,D) = 
\frac{L_\alpha}{4\pi D^2 T^4_\alpha F_3(\eta_\alpha)} \frac{E^2}{e^{E/
T_{\alpha}-\eta_\alpha}+1}~, 
\label{eq3} 
\end{eqnarray}
where $D$ is the distance to the supernova (typically $D\sim 10$ kpc for a
galactic supernova), $E$ is the energy of the neutrinos, $L_\alpha$ is the 
luminosity of the flavor $\nu_\alpha$, and $T_\alpha$ represents the
effective temperature of the $\nu_\alpha$ gas inside the neutrinosphere. 
%%%%
The typical values of the average neutrino energies produced by the supernova 
simulations are 
\begin{equation}
\langle E_e \rangle =10 -12 ~{\rm MeV}~, \hskip 0.4truecm  \langle
E_{\bar e} 
\rangle=12 - 18 ~{\rm MeV}~, \hskip 0.4truecm \langle E_x \rangle 
=18 - 27~{\rm MeV}~,
\label{temp}
\end{equation} 
and the integrated luminosities of all neutrino species are expected to 
be approximately equal, within a factor of two or so 
\cite{Raffelt:2002tu,Raffelt:1996wa}.
%%%%
%
The pinching parameter $\eta_\alpha$ takes the values  $\eta_\alpha \sim 0
- 2$ for $\numu$ and $\nutau$; larger values, $\eta_\alpha \sim 0 - 5$, could 
be realized for the electron flavor \cite{Janka,Raffelt:1996wa}. 
The normalization factor  $F_3(\eta_\alpha)$ is given by 
\begin{equation}
F_3(\eta_\alpha)\equiv -6 {\rm Li}_4(-e^{\eta_\alpha})~,
\label{f3}
\end{equation}
where the polylogarithm function ${\rm Li}_n(z)$ is defined as 
\begin{equation}
{\rm Li}_n(z)\equiv \sum_{k=1}^{\infty} \frac{z^k}{k^n}~.
\label{Li}
\end{equation}
In the absence of pinching, $\eta_\alpha=0$, one gets $F_3(0) = 
7\pi^4/120\simeq 5.68$.

\noindent
The average energy $\langle E_\alpha \rangle$ of the spectrum depends on both 
$T_\alpha$ and $\eta_\alpha$: 
\begin{eqnarray}
&&\langle E_\alpha \rangle=\sigma (\eta_\alpha) T_{\alpha}~,
\label{rel} \\
&&\sigma (\eta_\alpha) \equiv 3 \frac{{\rm Li}_4(-e^{\eta_\alpha})}
{{\rm Li}_3(-e^{\eta_\alpha})}~.
\label{sigma}
\end{eqnarray}
Eqs. (\ref{Li}) and (\ref{sigma}) give $\sigma(0)  = 3.15 $; larger values of 
$\sigma$ are obtained for larger (positive) $\eta_\alpha$, e.g. $\sigma(2) 
\simeq 3.6 $.

The luminosity, temperature and pinching  of the neutrino flux vary with  
the time $t$; if these variations occur over time scales larger than the duration 
of the burst, $\tau \sim 10$ s, the integrated $\nu_\alpha$ flux is still 
given by the expression (\ref{eq3}) with $\eta_\alpha \simeq 0$ and the total 
energy  $E_\alpha \sim  L_\alpha \tau $ in place of the luminosity $L_\alpha$. 
\\

In our study of neutrino conversions inside the star we use the following 
matter density profile of the star:
\begin{eqnarray}
\rho(r)=10^{13}~ C  \left(\frac{10 ~{\rm km}}{r} \right)^3~{\rm g\cdot 
cm^{-3}} ~, 
\label{eq4}
\end{eqnarray}
with $C\simeq 1 - 15$.  For $\rho \sim 1 - 10^{6} ~{\rm g\cdot
cm^{-3}}$ expression (\ref{eq4}) provides a good description of the
matter distribution at any time during the neutrino burst emission and
propagation through the star
\cite{BBB,Notzold:1987vc,Kuo:1988qu,Janka}. For $\rho \lta 1~{\rm
g\cdot cm^{-3}}$ the exact shape of the profile depends on the details
of evolution of the star, its chemical composition, rotation, etc.
The regions at $\rho \gta 10^{6}~{\rm g\cdot cm^{-3}}$ are reached by
the shock-wave propagation during the neutrino burst emission,
therefore in these regions the neutrinos cross the shock front, which
is characterized by large moving density gradients \cite{Fuller}.

%%%%%%%%%%%%%%%%%%%%%%%%%%%%%%%%%%%%%%%%%%%%%%%%%%%%%%%%%%%%%%%%%%%%%%%%%%%
\subsection{$\numu$ and $\nutau$ spectral difference}
\label{sect:2.2}

In the absence of specific predictions for the $\numu$ -- $\nutau$
flux differences, which could result from a number of physical
effects, we consider for these two species the Fermi-Dirac spectra, as
in eq. (\ref{eq3}), with different temperatures, luminosities and
pinching parameters.  Then the ratio of fluxes equals:
\begin{equation}
\frac{F^0_{\tau}}{F^0_{\mu}}= \left( \frac{L_{\tau}}{L_{\mu}} \right) 
\left(\frac{T_{\mu}}{T_{\tau}} \right)^4  \frac{F_3(\eta_\mu)}{F_3(\eta_\tau)} 
 \frac{\exp[E / T_\mu -\eta_\mu]+1}{\exp[E / T_\tau-\eta_\tau]+1} ~.
\label{rat}
\end{equation}
We consider the relative difference 
\begin{equation}
\Delta_F \equiv \frac{F^0_{\tau}}{F^0_{\mu}}-1~ 
\label{reldiff}
\end{equation}
and  adopt a general parameterization of $\Delta_F$ in terms of the 
relative differences of the temperatures, luminosities and pinching parameters: 
$\epsilon_T\equiv (T_\tau - T_\mu)/T_\mu$,  $\epsilon_L\equiv (L_\tau - L_\mu)
/L_\mu$ and $\epsilon_\eta\equiv (\eta_\tau - \eta_\mu)/\eta_\mu$.  The 
presence of charged muons in the star may result in a difference in the 
pinching parameters with no appreciable difference in the luminosities and 
temperatures, while different temperatures or luminosities  could appear 
from the effects of lepton number non-conservation or violation of lepton
universality.  

For $|\epsilon_T|,|\epsilon_L|,|\epsilon_\eta| \ll 1$ the difference 
$\Delta_F$ can be expanded as 
\begin{equation}
\Delta_F \simeq \epsilon_L + \epsilon_T \left[-4 + \frac{\exp[E / 
T_\mu-\eta_\mu]}{\exp[E / T_\mu-\eta_\mu]+1} \left( \frac{E}{T_\mu} \right) 
\right]+ \epsilon_\eta \eta_\mu \left[-\frac{{\rm Li}_3(-e^{\eta_\mu})}
{{\rm Li}_4(-e^{\eta_\mu})} + \frac{\exp[E / T_\mu-\eta_\mu]}{\exp[E / 
T_\mu-\eta_\mu]+1} \right] ~, 
\label{expa} 
\end{equation}
for any value of the energy $E$.
It is also useful to consider the high-energy limit, $E\gg T_\mu \eta_\mu$, of 
eqs. (\ref{rat}) and (\ref{expa}):
\begin{equation}
\frac{F^0_{\tau}}{F^0_{\mu}}(E\gg T_\mu \eta_\mu) \sim  \left(\frac{L_{\tau}}
{L_{\mu}} \right) \left(\frac{T_{\mu}}{T_{\tau}} \right)^4  
\frac{F_3(\eta_\mu)}{F_3(\eta_\tau)}  
 e^{ \eta_\tau -\eta_\mu} e^{ E (1/T_\mu -1/T_\tau)} ~,
\label{ratlim}
\end{equation}
and
\begin{equation}
\Delta_F (E\gg T_\mu \eta_\mu) \sim \epsilon_L + \epsilon_T \left[-4 + \left(
\frac{E}{T_\mu} \right) \right]+ \epsilon_\eta \eta_\mu 
\left[-\frac{{\rm Li}_3(-e^{\eta_\mu})}{{\rm Li}_4(-e^{\eta_\mu})} + 1  
\right] ~. 
\label{expalim} 
\end{equation} 

Let us comment on  the contributions of $\epsilon_T$, $\epsilon_L$ 
and $\epsilon_\eta$ to $\Delta_F$. 

1). The effect of different
luminosities, $\epsilon_L \neq 0$,  gives an energy-independent term in 
$\Delta_F$, as can be seen in eqs. (\ref{rat})-(\ref{expalim}). 
If there are no differences in the temperatures and in the pinching
factors one merely has $\Delta_F=\epsilon_L$.   
 
2). If the $\numu$ and $\nutau$ temperatures are different, 
$\epsilon_T\neq 0$, a 
critical energy $E^{corr}_C$ exists at which the $\numu$ and $\nutau$ fluxes 
are equal, $\Delta_F=0$.  
If $\nutau$ has a harder spectrum than $\numu$ ($\epsilon_T>0$) 
we have  $\Delta_F >0$ at  $E > E^{corr}_C$ and $\Delta_F < 0 $  for  
$E < E^{corr}_C$. For a given $\epsilon_\eta$, the critical energy 
$E^{corr}_C$  increases with the increase of  $\epsilon_T$ and decreases with 
the increase of $\epsilon_L$. If $\epsilon_T<0$ the opposite situation is 
realized: $\Delta_F >0$ at $E\lta  E^{corr}_C$ and $\Delta_F <0$ at $E\gta  
E^{corr}_C$.  The critical energy $E^{corr}_C$  increases with the increase of 
$\epsilon_T$ (the decrease in absolute value) and  with the increase of 
$\epsilon_L$. Above the energy  $E^{corr}_C$ the relative difference 
$\Delta_F$ of the fluxes increases linearly with $E$ in absolute value, as 
follows from eq. (\ref{expalim}). Numerically, taking $T_\mu=7$ MeV,
$\epsilon_T=0.1$ and $\epsilon_L=\epsilon_\eta=0$, we find $E^{corr}_C\simeq 
30$ MeV and eq. (\ref{expa})  gives $\Delta_F\simeq 0.46$ 
($\Delta_F\simeq 0.49$) at $E=60$ MeV and  $\Delta_F\simeq -0.35$ 
($\Delta_F\simeq -0.29$) at $E=5$ MeV. In general, for $\epsilon_T\neq 0$ the 
contribution of the originally non-electron neutrinos to the observed $\nue$ 
spectrum is dominated by the flavor with the softer spectrum in the low 
energy region, while at high energies the flavor with harder spectrum  
dominates.

3). The contribution due to different pinching parameters, $\epsilon_\eta 
\neq 0$, 
increases with energy and turns from negative to positive at $E\sim 
\langle E_\mu \rangle \sim 18  - 27 $ MeV.  At larger energies the 
$\epsilon_\eta $ contribution asymptotically approaches a constant value, as 
one can see from eqs. (\ref{ratlim}) and (\ref{expalim}). This value is small 
since the ratio ${{\rm Li}_3(-e^{\eta_\mu})}/{{\rm Li}_4(-e^{\eta_\mu})}$ is 
close to unity and  therefore it undergoes partial cancellation  in 
eq.  (\ref{expalim}). Taking  $\eta_\mu=1$, $\eta_\tau=2$ and $\epsilon_L=
\epsilon_T=0$ one gets  $\Delta_F\simeq 0.14$  at $E\gta 40$ MeV and  
$\Delta_F\simeq -0.46$  at $E=5$ MeV. 

The description given here applies to the difference of the $\barnumu$ 
and $\barnutau$ fluxes, $F^0_{\bar \mu}$ and  $F^0_{\bar \tau}$, as well. 
The relative difference $\bar \Delta_F\equiv F^0_{\bar \tau }/F^0_{\bar \mu}
-1$ can be described in terms of the ratios  $\bar \epsilon_T\equiv (T_{\bar 
\tau} - T_{\bar \mu})/T_{\bar \mu}$, $\bar \epsilon_L\equiv (L_{\bar \tau} - 
L_{\bar \mu})/L_{\bar \mu}$, and $\bar \epsilon_\eta\equiv (\eta_{\bar \tau} - 
\eta_{\bar \mu})/\eta_{\bar \mu}$ analogously to eq. (\ref{expa}).

%%%%%%%%%%%%%%%%%%%%%%%%%%%%%%%%%%%%%%%%%%%%%%%%%%%%%%%%%%%%%%%%%%%%%%%%%%%
%%%%%%%%%%%%%%%%%%%%%%%%%%%%%%%%%%%%%%%%%%%%%%%%%%%%%%%%%%%%%%%%%%%%%%%%%%%
\section{Conversion effects in the star}
\label{sect:3}
%%%%%%%%%%%%%%%%%%%%%%%%%%%%%%%%%%%%%%%%%%%%%%%%%%%%%%%%%%%%%%%%%%%%%%%%%%%

%%%%%%%%%%%%%%%%%%%%%%%%%%%%%%%%%%%%%%%%%%%%%%%%%%%%%%%%%%%%%%%%%%%%%%%%%%%
\subsection{Neutrino mass and mixing schemes}
\label{sect:3.1}

Let us consider a system of three neutrinos,  $(\nu_1,\nu_2,\nu_3)$, with 
masses $(m_1, m_2, m_3)$, related to the flavor eigenstates, $(\nue,\numu,
\nutau)$, by the mixing matrix $U$: $\nu_\alpha=\sum_{i} U_{\alpha i} \nu_i$. 
The indices $\alpha$ and $i$ refer to the flavor and mass eigenstates
respectively. 

We adopt  the following parameterization  of the mixing matrix (see e.g. 
\cite{Krastev:yu}):
\begin{equation} 
U=V^{23} I^{\delta} V^{13} V^{12} ,
\label{eq:U3nu}
\end{equation}
where 
\begin{equation}
I^{\delta}\equiv diag(1, 1 ,e^{i\delta} )~,
\label{idelta} 
\end{equation}
and
\begin{equation}
%{\footnotesize
V^{12}=\left(\begin{array}{ccc}
c_{12}    &  s_{12}  & 0     \\
-s_{12}   &  c_{12}    & 0     \\
0         &    0       & 1
\end{array} \right)\,,\quad
V^{13}=\left(\begin{array}{ccc}
c_{13}                  &    0       & s_{13} \\
0                       &    1       & 0       \\
-s_{13} &    0       & c_{13}
\end{array} \right)\, ,\quad
V^{23}=\left(\begin{array}{ccc}
1         &    0       & 0        \\
0         &  c_{23}    & s_{23}    \\
0         & -s_{23}   & c_{23}
\end{array} \right)\,,
%}
\label{1eq:V3}
\end{equation}
with $s_{ij} \equiv \sin \theta_{ij}$ and  $c_{ij} \equiv \cos \theta_{ij}$.

We will consider 3$\nu$-schemes which explain the atmospheric and the
solar neutrino data. The parameters describing the
atmospheric neutrino oscillations are \cite{Fukuda:2000np}: 
\begin{equation}
m_3^2 - m_2^2 \equiv    \Delta m^2_{32} = \pm (1.5 - 4) \cdot 10^{-3} 
{\rm eV}^2, ~~~~~~~ \tan^2 \theta_{23} = 0.48 - 2.1 ~.  
\label{atmpar}
\end{equation}
The two possibilities, $\Delta m^2_{32} >0$ and $\Delta m^2_{32} <0$, are 
referred to as {\it normal} and {\it inverted} mass hierarchies respectively.

The solar neutrino data  are explained either by vacuum oscillations (VO 
solution), or by one of the MSW solutions (LMA, SMA or LOW) with the
oscillation parameters 
\begin{equation}
m_2^2 - m_1^2 \equiv   \Delta m^2_{21} , ~~~~~~\tan^2 \theta_{12} ~,   
\label{sunpar}
\end{equation} 
(see, e.g., numerical values from the analyses
\cite{Fogli:2001vr}-\cite{Ahmad:2002ka}).

The angle $\theta_{13}$ is bounded to be small from the results of the CHOOZ 
and Palo Verde experiments \cite{Apollonio:1999ae,Boehm:2000vp}:
\begin{equation}
\sin^2 \theta_{13} \lta 0.02~, 
\label{ue3}
\end{equation}
and the CP-violating  phase $\delta$ is still completely undetermined.

%%%%%%%%%%%%%%%%%%%%%%%%%%%%%%%%%%%%%%%%%%%%%%%%%%%%%%%%%%%%%%%%%%%%%%%%%%%

\subsection{Matter effects: $V_{\tau \mu}$ potential; the level crossing
pattern}
\label{sect:extra}

Due to the very wide range of matter densities inside a supernova, $\rho \sim  
0 - 10^{13}~{\rm g\cdot cm^{-3}}$, the conditions for three MSW resonances
(level crossings) are met. 
The crossing of  $\numu$ and $\nutau$ levels, 
the $\mu\tau$ resonance,  
is driven by the difference of the $\numu$ and $\nutau$ potentials  
which appears at one loop level due to different  
masses of the muon and tau leptons~\cite{Botella:1987wy}:
\begin{equation}
V_{\tau \mu} \equiv V_\tau - V_\mu \simeq \pm \frac{3}{2 \pi^2} G_F^2
m_\tau^2 
\left[(n_p + n_n)\ln\left( \frac{M_W}{m_\tau}\right) -n_p - \frac{2}{3} n_n 
\right] ~,
\label{1pot}
\end{equation}
where $G_F$ is the Fermi constant, $ m_\tau$ and $M_W$ are the masses of the 
tau lepton and of the W boson respectively; $n_p$ and $n_n$ denote the number 
densities of protons and neutrons in the medium.  In eq. (\ref{1pot}) the $+$ 
($-$) sign refers to neutrinos (antineutrinos).
The  difference $V_{\tau \mu} $  
is much smaller than the $\nue - \numu/\nutau$ effective 
potential $V_e =\sqrt{2} G_F n_e$, where $n_e$ is the electron number 
density of 
the medium.  One finds  
\begin{equation}
V_{\tau \mu} \simeq 5 \cdot 10^{-5} V_e ~.
\label{comp}
\end{equation} 
 
The conversion induced by the potential (\ref{1pot}) is governed by
the ``atmospheric'' mixing and mass splitting, $\Delta m^2_{32}$ and
$\theta_{23}$, eq. (\ref{atmpar}).  Depending on the precise value of
$\theta_{32}$ the level crossing occurs in the range of densities
$\rho \sim 0 - \rho_{\mu\tau}$, where $\rho_{\mu\tau}$ is determined
by the condition
\begin{equation} 
V_{\tau \mu}(r)\sim \frac{\Delta m^2_{32}}{2 E}~.
\label{mva}
\end{equation}
Numerically, one gets $ \rho_{\mu\tau} \sim  10^{7} -
10^{8} ~{\rm g\cdot cm^{-3}}$. In fact, as we will see,  
the actual position of the level crossing   is not important. 

The probability $P_{23}$ of the transition  between the eigenstates of the 
Hamiltonian in matter, $\nu_{3m}$ and $\nu_{2 m}$, can be calculated 
(following, e.g.,   \cite{Fogli:2001pm}) according to
\begin{eqnarray}
&&P_{23} = \frac{e^{\chi \cos^2 \theta_{23}} -1}{e^{\chi} -1}~,
\label{adiab} \\
&&\chi \equiv - 2 \pi \frac{\Delta m^2_{32}}{2 E} \left[\frac{1}{V_{\tau \mu}
(r)} \frac{dV_{\tau \mu}(r)}{dr}\right]^{-1}_{r=r_p}~,
\label{chi}
\end{eqnarray}
where $r_p$ is 
the distance from the center of the star to the region defined by 
eq. (\ref{mva}).
 
With the profile (\ref{eq4}) and the parameters (\ref{atmpar}), eqs. 
(\ref{1pot}), (\ref{chi}) and (\ref{mva}) give $\chi= 500 - 900$, 
corresponding to a very good adiabaticity: $P_{2 3}=0$. 
The adiabaticity is good also for steeper profiles. We have checked 
that the adiabaticity condition is satisfied even for the  large
density gradients which could be produced by the shock wave 
propagation. 
\\

The two other resonances, driven by the $\nue - \numu/\nutau$
effective potential, $V_e$, occur at much smaller densities, $\rho
\lta 10^{4} ~{\rm g\cdot cm^{-3}}$.  For these values of the density
the potential $V_{\tau \mu}$ is negligibly small and it is convenient
to consider the evolution of the system in the rotated basis of
states:
\begin{equation}
\nu_\alpha'\equiv I^{\delta \dagger} V^{{23} \dagger} \nu_\alpha~.
\label{rotstat}
\end{equation}
\\

\noindent
The second (H-)  resonance, associated 
to  $\Delta m^2_{32}$ and $\theta_{13}$,   occurs at   
$\rho = \rho_H \sim 10^3 - 10^4 ~{\rm g  \cdot cm^{-3}}$.  
The resonance  is in the neutrino (antineutrino) 
channel if the mass hierarchy is normal (inverted). We denote as $P_H$ the
probability of the transition between the eigenstates of the Hamiltonian in 
this resonance (the hopping probability). As shown in refs. 
\cite{Dighe:1999bi,Lunardini:2001pb}, $P_H$ increases with the decrease of $\sin^2 
\theta_{13}$, varying from perfectly adiabatic conversion ($P_H=0$) at 
$\sin^2 \theta_{13}\gta 10^{-3}$ to strong adiabaticity breaking ($P_H=1$) at 
$\sin^2 \theta_{13}\lta 10^{-5}$. 

The third level crossing (L resonance) takes place at a lower density,
$\rho=\rho_L \lta 10 ~{\rm g \cdot cm^{-3}}$. It is determined by the
``solar'' parameters $\Delta m^2_{21}$ and $\theta_{12}$,
eq. (\ref{sunpar}), and occurs in the neutrino channel (see sec.
\ref{sect:3.1}).
We denote the hopping probability associated to this resonance as $P_L$.  
For oscillation parameters in the LMA region the conversion is adiabatic, 
$P_L=0$, while partial breaking of adiabaticity occurs for other 
solutions of the solar neutrino problem \cite{Dighe:1999bi,Lunardini:2001pb}. 
\\

The level crossing scheme for the normal mass hierarchy with 
$\theta_{23}<\pi/4$ 
and a large solar mixing angle $\theta_{12}$ is given in fig. 
\ref{fig:levcross}. 
The dynamics of neutrino conversions inside the star can be considered 
as occurring in two steps: 

(1) propagation between the neutrinosphere and the layer with an intermediate 
density $\rho_{int}$ such that $\rho_H <  \rho_{int} \ll \rho_{\mu \tau}$, 
(``inner region") and
 
(2) propagation from the layer with $\rho\sim \rho_{int}$ to the
surface of the star (``outer region'').

Let us consider the evolution in the inner region. At the
neutrinosphere the potentials $V_e$ and $V_{\tau\mu}$ dominate over
the kinetic energy differences $\Delta m_{ij}^2/2E$ and the mixing is
strongly suppressed, i.e. the
matter eigenstates coincide with the flavor ones. At the densities
$\rho\sim \rho_{int}$, the potential $V_{\tau\mu}$ is negligible and
the matter eigenstates coincide with the states of the rotated basis
(\ref{rotstat}).  Using the level crossing scheme of fig. 1 and taking
into account that the adiabaticity is satisfied in whole range of
densities $\rho_H < \rho_{int} \ll \rho_{\mu \tau}$, we find that in
the inner region the following transitions take place:
\begin{eqnarray}
\nue\to\nue\,,\quad\quad\quad \numu\to\numu'\,,\quad\quad\quad\nutau\to
\nutau'\,, \nonumber \\
\barnue\to\barnue\,,\quad\quad\quad \barnumu\to\barnutau '\,,\quad\quad\quad
\barnutau\to \barnumu '\,,
\label{transit}
\end{eqnarray}
provided that the $\mu\tau$ level crossing occurs in the antineutrino channel. 
Therefore in the rotated basis $(\nue, \numu', \nutau')$ the fluxes of the 
states (before they enter the H resonance region) are 
$$ 
F(\nue) =  F_e^0, ~~~  F(\numu') =  F_\mu^0, ~~~  F(\nutau') = F_\tau^0~.
$$  
Actually, the observables  do not depend on the position of the $\mu\tau$  
resonance: what matters is the flux of the state that 
crosses the $\nue$ level at the H resonance.  

Notice that the role of the potential $V_{\tau\mu}$ is reduced to the 
suppression of the flavor mixing at high densities; the adiabatic
change of this potential allows one to relate the fluxes of the states of
the rotated basis to those of the originally produced flavor states. 

The evolution of the neutrino states in the outer region is determined by 
the level crossings at the H and L resonances. 
If the hierarchy is inverted, then, as shown in fig. \ref{fig:levcross2}, the
$\mu\tau$ and L resonances are in the neutrino channel, while the H resonance 
is in the antineutrino channel. In the inner region the conversions  
$\nutau \rightarrow \numu'$ and $\numu \rightarrow \nutau'$ take place,  
while the $\nue \leftrightarrow \numu'$ conversion occurs in the L
resonance. Antineutrinos first undergo the $\barnumu \rightarrow \barnumu
'$ and $\barnutau \rightarrow \barnutau '$ conversions in the inner 
region. At lower densities, the $\barnue \leftrightarrow \barnumu' /
\barnutau '$ transitions take place in the H resonance and in the L resonance 
for large $\theta_{12}$.

%%%%%%%%%%%%%%%%%%%%%%%%%%%%%%%%%%%%%%%%%%%%%%%%%%%%%%%%%%%%%%%%%%%%%%%%%%%
\subsection{Conversion probabilities}
\label{sect:3.2}

As discussed in \cite{Dighe:1999bi}, if $F^0_\mu=F^0_\tau\equiv F^0_x$ the 
effect of the conversion on the flux  $F_{ e}$  of electron neutrinos
observed at the Earth can be expressed in terms of a single permutation
parameter, $p$, which represents the $\nue$ survival probability:   
\begin{equation}
F_{ e}=p F^0_{ e}+(1-p) F^0_x~.
\label{2case}
\end{equation}

\noindent
We shall generalize now this result to the case $F^0_\mu \neq F^0_\tau$.  
\\

Let us show that the differences between the $\numu$ and $\nutau$ ($\barnumu$ 
and $\barnutau$) fluxes can manifest themselves only in the following
cases:  
\\

\noindent
(i) normal hierarchy, neutrino channel,  
\\

\begin{figure}[hbt]
\begin{center}
\psfrag{r}{$n_e$}
\psfrag{a}{$\nue$}
\psfrag{b}{$\nutau$}
\psfrag{c}{$\numu$}
\psfrag{d}{$\nu_{3 m}$}
\psfrag{e}{$\nu_{2 m}$}
\psfrag{f}{$\nu_{1 m}$}
\psfrag{g}{$\numu'$}
\psfrag{h}{$\nutau'$}
\psfrag{j}{$\barnumu$}
\psfrag{k}{$\barnutau$}
\psfrag{w}{$\barnue$}
\psfrag{X}{$\mu \tau$}
\epsfig{file=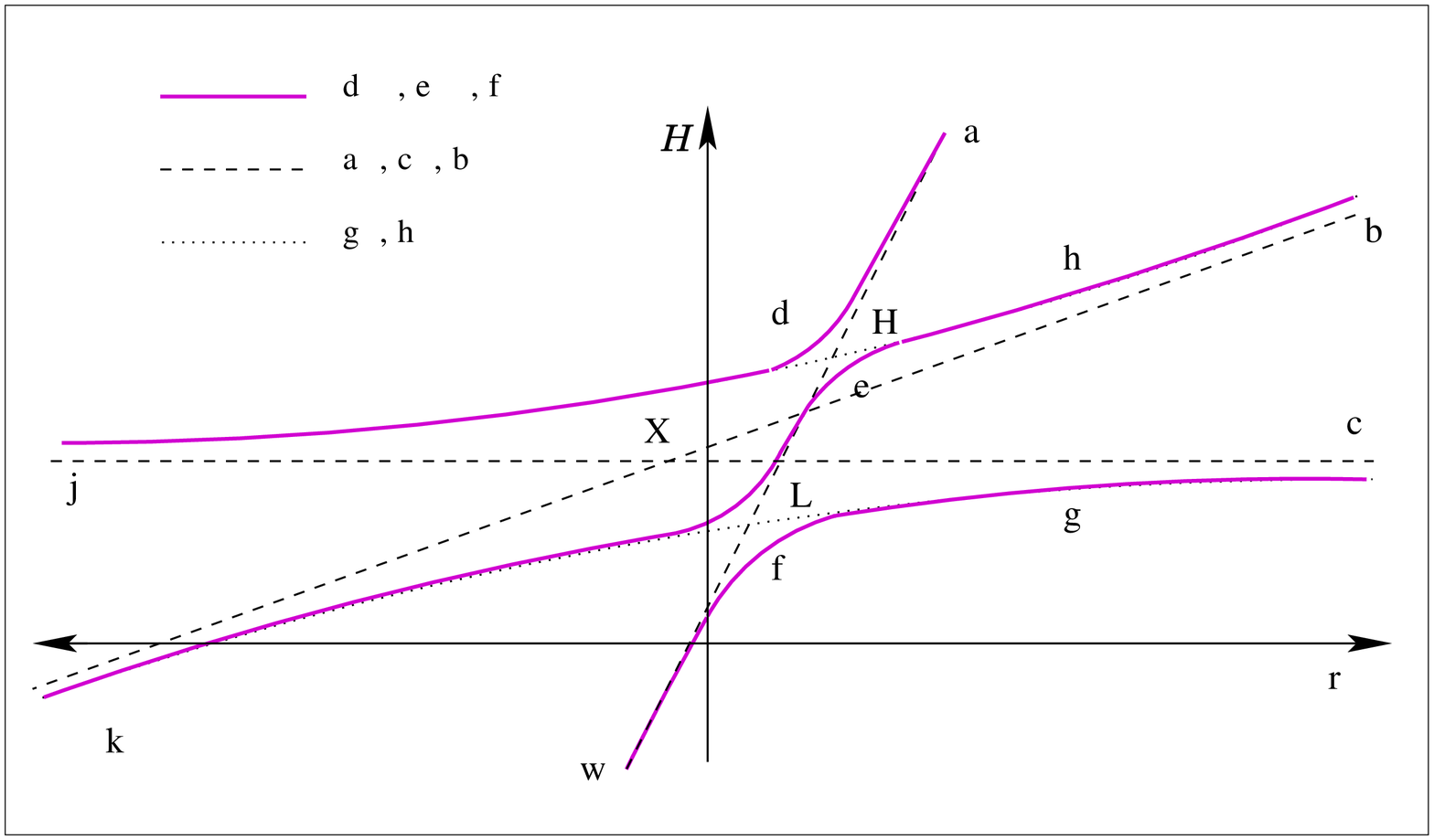, width=17truecm}
\end{center}
\caption{The level crossing scheme of neutrino conversion in the matter of
the star. The figure refers to the case of normal mass hierarchy with
$\theta_{23}< \pi/4$ and a large solar angle $\theta_{12}$. The semi-plane
with negative density, $n_e<0$, describes the conversion in the
antineutrino channel.} 
\label{fig:levcross}
\end{figure}

\begin{figure}[hbt]
\begin{center}
\psfrag{r}{$n_e$} 
\psfrag{a}{$\nue$} 
\psfrag{b}{$\nutau$}
\psfrag{c}{$\numu$} 
\psfrag{d}{$\nu_{3 m}$} 
\psfrag{e}{$\nu_{2 m}$}
\psfrag{f}{$\nu_{1 m}$} 
\psfrag{g}{$\numu'$} 
\psfrag{h}{$\nutau'$}
\psfrag{j}{$\barnumu$} 
\psfrag{k}{$\barnutau$} 
\psfrag{w}{$\barnue$}
\psfrag{X}{$\mu \tau$} 
\epsfig{file=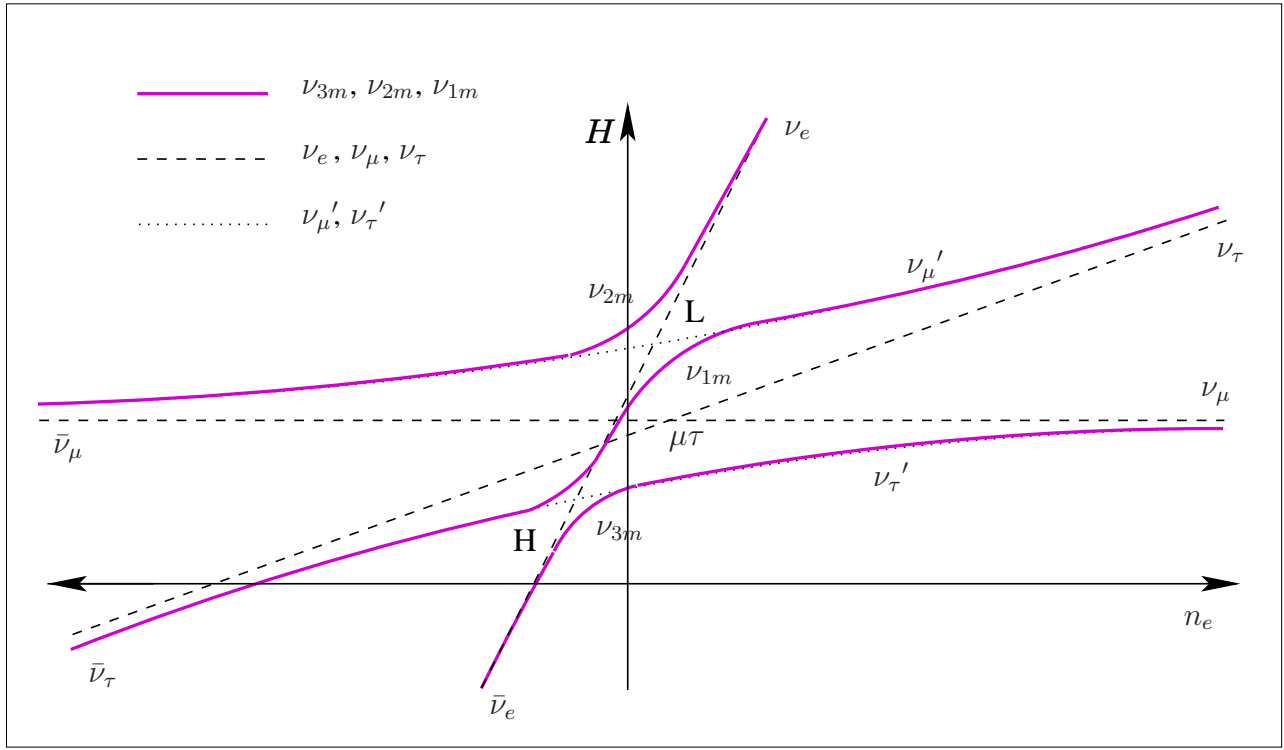, width=17truecm}
\end{center}
\caption{The same as in fig. \ref{fig:levcross} but for the inverted mass
hierarchy.}
\label{fig:levcross2}
\end{figure}

\noindent
(ii) inverted hierarchy, antineutrino channel,   
\\

\noindent
both of them with non-maximal adiabaticity breaking in the H resonance: 
$P_H<1$. Indeed, as can be seen from figs. \ref{fig:levcross} and
\ref{fig:levcross2}, in the remaining cases (inverted hierarchy, neutrino
channel and normal hierarchy, antineutrino channel) neutrinos of one of
the non-electron flavors are completely converted into the third  mass
eigenstate, for which the electron component is small (equal to
$\sin^2 \theta_{13}$). Therefore this  flavor does not contribute 
significantly to the observed $\nue$ (or  $\barnue$) flux, making the
study of the $\numu$ -- $\nutau$  inequivalence impossible. The same  
conclusion  is obtained in the schemes (i) and (ii) for maximal  
adiabaticity  breaking in the H resonance, $P_H=1$, since this corresponds
to the complete transition of one of the non-electron flavors to the third
mass eigenstate.

In what follows  we discuss the conversion probabilities for the scenarios
(i) and (ii); results  for the remaining cases can be obtained by a simple
generalization.\\

Let us first consider  the conversion of neutrinos in the scheme with
normal hierarchy, $\Delta m^2_{32}>0$.
As we discussed in sec.  \ref{sect:extra}, the neutrinos arrive at the
region of the H resonance as incoherent states $\nue$, $\numu'$,$\nutau'$ 
with the fluxes $F_e^0$, $F_\mu^0$, $F_\tau^0$. Let us consider the
further evolution of these states. As a result of the adiabatic
conversions in the star and/or decoherence due to the spread of the
wavepackets, the neutrinos leave the star as mass eigenstates.  The fluxes
${\vec {\cal F}}\equiv \left(F_1,F_2, F_3 \right)$ of these states  can be
expressed in terms of the original flavor fluxes ${\vec F^0}\equiv 
\left(F^0_e,F^0_{\mu}, F^0_\tau \right)$ according to 
\begin{equation}
{\vec {\cal F}}={\cal P} {\vec F^{0}}~,
\label{convmat}
\end{equation}   
where ${\cal P}$ is the  matrix of the conversion probabilities inside the
star in the outer region, ${\cal P}_{i \alpha}\equiv P({\nu_{\alpha}
\rightarrow \nu_i})$. In the approximation of factorized dynamics,
(factorization of the transition probabilities in H- and L- resonances)
 the matrix  ${\cal P}$ is \cite{Dighe:1999bi}
\begin{equation}
{\cal P}=\left(\begin{array}{ccc}
P_H P_L         &    1-P_L       & (1-P_H) P_L       \\
P_H (1-P_L)       &  P_L    &  (1-P_H) (1-P_L) \\
 (1-P_H)        & 0    & P_H
\end{array} \right)\,,
\label{calp}
\end{equation} 
as can be readily derived from the level crossing scheme, fig. 
\ref{fig:levcross}. 
Then the  fluxes of neutrinos of definite flavors observed at the Earth,
${\vec F}\equiv \left(F_e,F_{\mu}, F_\tau \right)$, are obtained by
the projection
\begin{equation}
{\vec F}=S {\vec {\cal F}}~,
\label{proj}
\end{equation} 
where the matrix $S$ is defined as 
\begin{equation}
S_{\alpha i}\equiv |U_{\alpha i}|^2~.
\label{smatrix}
\end{equation}
Combining eqs. (\ref{convmat})-(\ref{smatrix}) one finds the relation
between the original fluxes and the observed flavor fluxes:
\begin{eqnarray}
&&{\vec F}=Q {\vec {F^0 }}~, 
\label{final}\\
&&Q\equiv S {\cal P}~.
\label{Pmatrix}
\end{eqnarray} 
Each entry $Q_{ \beta \alpha}$ of the matrix $Q$  represents the total
$\nu_\alpha \rightarrow \nu_\beta$ conversion probability, i.e. the
probability that a neutrino produced as $\nu_\alpha$ in the star is
observed as $\nu_\beta$ in the detector. From eqs. (\ref{eq:U3nu}), 
(\ref{1eq:V3}), (\ref{smatrix}), (\ref{Pmatrix}) and (\ref{calp}),
neglecting terms of order  $\sin^2 \theta_{13}$, we get   
\begin{equation}
Q=Q^{(0)}+2\tilde{J}Q^{(1)}~,
\label{qexp}
\end{equation}
where 
\begin{equation}
\tilde{J}\equiv s_{12} c_{12} s_{23} c_{23} s_{13} \cos \delta\,,
\label{tildej}
\end{equation}
and
\begin{eqnarray}
&&Q^{(0)} = \left(\begin{array}{ccc}
P_H f_L      &  1 -f_L    & (1-P_H) f_L     \\
c^2_{23} P_H (1-f_L)+s^2_{23} (1-P_H) & c^2_{23} f_L  &
c^2_{23}(1-P_H)   (1-f_L)+s^2_{23}  P_H   \\
s^2_{23} P_H (1-f_L)+c^2_{23} (1-P_H) & s^2_{23} f_L  &
s^2_{23}(1-P_H)  (1-f_L)+c^2_{23}  P_H  
\end{array} \right)\,, \nonumber
\\
\label{Q0}\\
&&Q^{(1)} =(1-2P_L)\left(\begin{array}{ccc}
  0   & 0  &  0   \\
 -P_H    & 1  & - (1-P_H)     \\
  P_H    & -1  & (1-P_H)
\end{array} \right)\,.
\label{Q1}
\end{eqnarray} 
In eqs. (\ref{Q0})
\begin{equation}
f_L\equiv P_L + (1-2 P_L) \sin^2 \theta_{12}~. 
\label{fl}
\end{equation}
As can be seen from eq. (\ref{Q1}), the observed $\nue$
flux does not depend on the CP-violating phase $\delta$ and, moreover,
the terms proportional to $\tilde{J}$ cancel once the sum
$F_\mu+F_\tau$ is considered.  Therefore, since the separate detection
of $\numu$ and $\nutau$ is impossible, 
the study of the CP-violating phase $\delta$ with supernova neutrinos
seems to be not feasible.  This result has a general validity and is
discussed in more detail in sec.  \ref{sect:cpviol}.
\\

We consider now the  antineutrino channel in the case of the inverted 
hierarchy, $\Delta m^2_{32}<0$ (fig. \ref{fig:levcross2}). 
Similarly to eqs. (\ref{final})-(\ref{Pmatrix}), the fluxes $ {\vec {\bar
F}}\equiv \left(F_{\bar e},F_{\bar \mu}, F_{\bar \tau} \right)$ of the
antineutrinos at the detector are related to the original fluxes $ {\vec 
{\bar F}^0}\equiv \left(F^0_{\bar e},F^0_{\bar \mu}, 
F^0_{\bar \tau}  \right)$  by 
\begin{eqnarray}
&&{\vec {\bar F}}={\bar Q} {\vec {\bar F}^0 }~, 
\label{finalb}\\
&&\bar Q\equiv S \bar{\cal P}~.
\label{Pmatrixb}
\end{eqnarray} 
The matrix  $\bar {\cal P}$  of the ${\bar \nu}_\alpha \rightarrow {\bar
\nu}_i$ conversion probabilities  
is given by 
\begin{equation}
\bar {\cal P}=\left(\begin{array}{ccc}
P_H (1-\bar P_L)         &   \bar P_L       & (1-P_H)(1-\bar P_L) \\
P_H   \bar P_L      &  (1-\bar P_L)   &  (1-P_H)\bar P_L \\
 (1-P_H)        & 0    & P_H
\end{array} \right)\,,
\label{xcalp}
\end{equation} 
where $\bar P_L$ represents the hopping probability 
associated to the L resonance in the antineutrino channel \cite{Dighe:1999bi}. 
A form analogous to eq. (\ref{qexp}) is obtained for the $\bar Q$ matrix,
$\bar Q=\bar Q^{(0)}+2\tilde{J} \bar Q^{(1)}$, where the terms $\bar Q^{(0)}$ 
and  $\bar Q^{(1)}$ are 
\begin{eqnarray}
&&\bar Q^{(0)} = \left(\begin{array}{ccc}
P_H {\bar f}_L      &  1 -{\bar f}_L    & (1-P_H) {\bar f}_L     \\
c^2_{23} P_H (1-{\bar f}_L)+s^2_{23} (1-P_H)     & c^2_{23} {\bar f}_L &
c^2_{23}(1-P_H)  (1-{\bar f}_L)+s^2_{23}  P_H   \\
s^2_{23} P_H (1-{\bar f}_L)+c^2_{23} (1-P_H) & s^2_{23} {\bar f}_L     &
s^2_{23}(1-P_H)  (1-{\bar f}_L)+c^2_{23}  P_H  
\end{array} \right)\,, \nonumber
\\
\label{xQ0}\\
&&\bar Q^{(1)} =-(1-2{\bar P}_L)\left(\begin{array}{ccc}
  0   & 0  &  0   \\
 -P_H    & 1  & - (1-P_H)     \\
  P_H    & -1  & (1-P_H)
\end{array} \right)\,.
\label{xQ1}
\end{eqnarray}  
Here we defined
\begin{equation}
\bar f_L\equiv \bar P_L + (1-2 \bar P_L) \cos^2 \theta_{12}~.
\label{flanti}
\end{equation}
One can see that the results (\ref{xcalp})-(\ref{xQ1}) can be obtained
from those in eqs. (\ref{calp})-(\ref{Q1}) by the replacement 
\begin{equation}
P_L \rightarrow 1-{\bar P}_L~,
\label{plrep}
\end{equation}
or, equivalently,
\begin{equation}
f_L \rightarrow \bar f_L.
\label{flrep}
\end{equation}

%%%%%%%%%%%%%%%%%%%%%%%%%%%%%%%%%%%%%%%%%%%%%%%%%%%%%%%%%%%%%%%%%%%%%%%%%%%
\subsection{Conversion effects and CP violation}
\label{sect:cpviol}

As follows from eqs. (\ref{Q1}) and (\ref{xQ1}), the elements of the
second and the third lines of the matrices $Q^{(1)}$ and
$\bar{Q}^{(1)}$ have opposite signs and therefore the sums
$F_\mu+F_\tau$ and $F_{\bar \mu}+F_{\bar \tau}$ of the observed
non-electron fluxes do not depend on $\delta$. Let us show that this
result holds in the general case, without any assumption about the
dynamics of neutrino conversion.  \\

\noindent
Consider the evolution of the neutrino states at densities $\rho \lta
\rho_{int} $, where the potential $V_{\tau \mu}$ can be neglected and
the conversion can be described in terms of the rotated states
(\ref{rotstat}).  The fluxes of these states coincide with the
original $\nu_\alpha$ fluxes as a result of the adiabatic conversion
in the $\mu\tau$ resonance. From the definition (\ref{rotstat}) it 
follows that the evolution of the rotated states $\nu_\alpha '$ and
therefore the 
matrix $\cal{P}$, eq. (\ref{calp}), do not depend on the CP-violating
phase $\delta$, irrespective of whether the approximation of
factorization of transitions in different resonances is used.  Terms
containing $\delta$ may appear in the conversion probabilities only
due to the projection (\ref{proj}) of mass states to the flavor basis
by the matrix $S$, eq. (\ref{smatrix}). We can write the sum of the
$\numu$ and $\nutau$ fluxes as
\begin{equation}
F_\mu+F_\tau=\sum\limits_{i \alpha}(S_{\mu i}+S_{\tau i})P_{i\alpha} 
F_\alpha^0\,.
\label{sum1}
\end{equation}
As a consequence of unitarity we find that for the $\delta$-dependent parts  
$S^{(\delta)}_{\alpha i}$ of the matrix $S$ the following relation holds: 
\begin{equation}
S^{(\delta)}_{e i }+S^{(\delta)}_{\mu i}+S^{(\delta)}_{\tau i}=0~;
\label{sum}
\end{equation}  
furthermore, in the parameterization (\ref{eq:U3nu}) we have:
\begin{equation}
S^{(\delta)}_{e i }=0~.
\label{sume}
\end{equation}
The combination of eqs. (\ref{sum}) and (\ref{sume}) yields 
\begin{equation}
S^{(\delta)}_{\mu i}+S^{(\delta)}_{\tau i}=0~.
\label{finalsum}
\end{equation}
This means that the sum (\ref{sum1}) of the $\numu$ and $\nutau$
 fluxes is independent of $\delta$.

Neutrino propagation in an asymmetric matter can give rise to
matter-induced T violation effects in neutrino oscillations even in the 
absence of the fundamental CP and T violation (i.e. when $\delta=0$). 
In \cite{AHLO} it was pointed out that such effects cannot be observed 
in supernova neutrino experiments as a consequence of the assumption 
$F_\mu^0=F_\tau^0$ and experimental indistinguishability of $\numu$ and
$\nutau$. We note here that this result holds true even if the 
condition $F_\mu^0=F_\tau^0$ is relaxed. This is a consequence of the
fact that the effects of matter-induced T violation (as well as those of 
the fundamental T violation) drop out from the sum (\ref{sum1}), irrespective 
of whether the original $\numu$ and $\nutau$ fluxes coincide. 

%%&&&&&&&&&&&&&&&&

%%%%%%%%%%%%%%%%%%%%%%%%%%%%%%%%%%%%%%%%%%%%%%%%%%%%%%%%%%%%%%%%%%%%%%%%%%%
\subsection{Expected spectra}
\label{sect:3.3}

Let us  discuss the observed $\nue$ signal for normal mass hierarchy (the
case (i) of sec. \ref{sect:3.2}).  

{}From  eqs. (\ref{final})-(\ref{Q1}) one finds that the detected flux of
electron neutrinos, $F_e$, is  
\begin{eqnarray}
F_{ e} &=& F_{ e}^0 Q_{e e}  +  F_{ \mu}^0 Q_{ e \mu}+ F_{ \tau}^0 Q_{ e
\tau}~, 
\label{decflux}\\
&=&F_{ e}^0 P_H  f_L  +  F_{ \mu}^0 (1-f_L) + F_{ \tau}^0 (1-P_H) f_L~.
\label{explicit}
\end{eqnarray}
These expressions reduce to the equal fluxes form, eq. (\ref{2case}), if
one of the following possibilities is realized: 

%\begin{itemize}
%\item  

%\item 
(1) the adiabaticity  is maximally violated in the H resonance, $P_H=1$
(as discussed in sec. \ref{sect:3.2}), 

%\item 
(2)$f_L=0$, or 
 
%\item 
(3) $f_L=1$. \\
%\end{itemize} 

\noindent
As can be seen from eq. (\ref{fl}), the last two possibilities are
realized only if the angle $\theta_{12}$ is small, which is disfavored
by the solar neutrino data (see e.g.
\cite{Fogli:2001vr}-\cite{Ahmad:2002ka}). In the following we focus
on large solar mixing angles and therefore neglect cases (2) and (3).

Expression (\ref{decflux}) can be conveniently rewritten as
\begin{equation}
F_{ e} = F_{ e}^0 P_H f_L  +  F_{ \mu}^0 (1-P_H f_L) + (F_{ \tau}^0-F_{
\mu}^0)(1-P_H) f_L  ~,
\label{rewr}
\end{equation}
where the third term is entirely due to the difference of the $\numu$ and
$\nutau$ fluxes and vanishes if either condition (1) or (2) is 
fulfilled.  
{}From eq. (\ref{rewr}) one finds the relative deviation, $R$, of the
observed $\nue$ flux from the one in the case of equal fluxes:
\begin{equation}
R = \Delta_F \frac{(1-P_H) f_L}{( F_{ e}^0/F_{ \mu}^0) P_H  f_L +   (1-P_H
f_L) } ~,
\label{reldev}
\end{equation}
where $\Delta_F$ was given in eqs. (\ref{reldiff}) and (\ref{expa}).
As follows from eq.  (\ref{reldev}), the sign of $R$ is determined  by 
that of 
$\Delta_F$. Therefore  $R$ vanishes at  the critical energy $E^{corr}_C$
and, if $\epsilon_T>0$  ($\epsilon_T<0$), we have $R< 0$ ($R> 0$)  for
$E<E^{corr}_C$ and $R> 0$  ($R< 0$) for $E>E^{corr}_C$.  As discussed in
sec. \ref{sect:2.2}, if  $\epsilon_T=\epsilon_\eta=0$ no change of sign
occurs and the quantity $R$ is determined by the difference in the
luminosities only.   

It is useful to introduce another critical energy $E^{(2)}_C$ defined as
the energy at which the difference $F_{ e}^0 - F_{ \mu}^0$ changes its sign
\cite{Dighe:1999bi}.  At $E \ll E^{(2)}_C$ the electron neutrino flux
dominates over the  $\numu$ and $\nutau$ fluxes; in contrast, for  $E \gg
E^{(2)}_C$ the latter  fluxes -- having larger average energy -- are
dominant.  Numerically,  taking $T_e=3.5$ MeV, $T_\mu=7$ MeV, $L_e=L_\mu$
and $\eta_e=\eta_\mu=0$ one finds $E^{(2)}_C \simeq 20$ MeV.

Taking $P_L=0$ (which is the case for solar neutrino
parameters in the LMA region), from eq. (\ref{reldev}) we find that 
\begin{equation} 
0 \leq \frac{R}{\Delta_F} \leq \sin^2 \theta_{12}~,
\label{bounds}
\end{equation}
for any value of the energy $E$.  The upper bound is reached if $P_H=0$
and $E\gg E^{(2)}_C$, for which the factor $F_{ e}^0/F_{ \mu}^0$ in
eq. (\ref{reldev}) is negligibly small due to the hierarchy of the 
temperatures. Notice that the condition $P_H=0$  corresponds 
to no contribution of the original $\nue$ flux to the observed $\nue$ 
signal: $Q_{ee}=0$ (see eq. (\ref{Q0})).  Thus, the observed $\nue$ flux
is given by a mixture of the original $\numu$ and $\nutau$ fluxes.
The effect is suppressed, $R\simeq 0$, for $P_H \sim 1$ or at low energies,
$E\ll E^{(2)}_C$, where the ratio $F_{ e}^0/F_{ \mu}^0$ is large. 
Numerically, taking $P_L=P_H=0$, $T_\mu=7$ MeV,
$\epsilon_T=0.1$,  $\eta_\mu=\eta_\tau=\epsilon_L=0$ and $\tan^2
\theta_{12} \simeq 0.33$ one gets $R \simeq -0.1$ at $E=5$ MeV and $R
\simeq 0.12$ at $E=60$ MeV. 

In general  for $|\epsilon_T|,|\epsilon_L|<0.2$ 
and $|\epsilon_\eta| \lta 1$ the relative deviation $R$ 
%of the observed  $\nue$
%spectrum from the one expected in the equal fluxes case, 
%$F_{\mu}^0= F_{ \tau}^0$, 
can be as large as $\sim 40 -50\%$ (in absolute value) at
high energies, $E \gg E^{(2)}_C$, and $\sim 15\%$ in the low energy
limit, $E \ll E^{(2)}_C$.

For antineutrino conversion  with  inverted mass hierarchy
the effects of the difference of the $\barnumu$ and $\barnutau$ fluxes are
described by a straightforward generalization of eqs. (\ref{decflux}) -
(\ref{reldev}). The results are similar to those presented here. \\

Let us consider the possibility of observation of the difference of the
$\numu$ and  $\nutau$  fluxes. It is easy to see that, despite relatively
large values of $R$ found here, the effect of the $\nu_{\mu}$ -- $\nu_{\tau}$ 
inequivalence will be very difficult to identify. Indeed, the corresponding 
distortion of the observed $\nue$ spectrum, which for $\epsilon_T\neq 0$
consists in a broadening of the energy distribution, can be mimicked by  
other effects and therefore does not represent a unique signature of
the unequal $\nu_{\mu}$ and $ \nu_{\tau}$ fluxes.

\begin{figure}[hbt]
\begin{center}
\epsfig{file=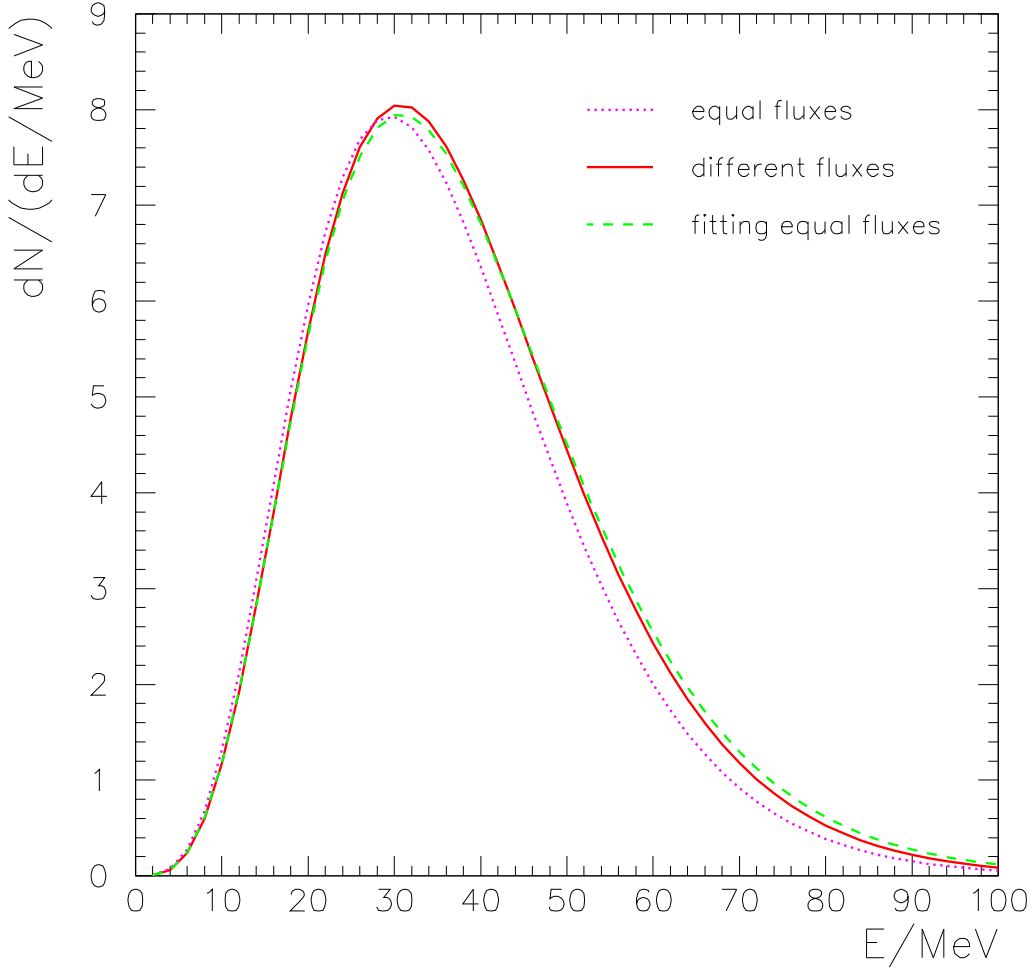, width=15truecm}
\end{center}
\caption{The  energy spectrum of the  $\nue + d \rightarrow e^- + p + p$
events expected at SNO for conversion in the star only with equal (dotted
and dashed curves) and different $\numu$ and $\nutau$ fluxes (solid
curve).  The dotted curve was obtained considering normal mass hierarchy
with  $P_H=0$,  $T_\mu=7$ MeV, $\eta_\mu=0$, and LMA oscillation
parameters ($P_L=0$ and $\sin^2 2\theta_{12}=0.75$). The solid curve
refers to the same parameters with the difference $\epsilon_T=0.2$. The
dashed curve mimicks the effect of the different $\numu$--$\nutau$ fluxes
with  $T_\mu=T_\tau=7.3$ MeV.  A distance $D=8.5$ kpc from the supernova
and integrated luminosity $E_\mu=E_\tau=5\cdot 10^{52}~{\rm ergs}$ were taken.}
\label{fig:f6} 
\end{figure}

In particular, taking, e.g., $\epsilon_L=\epsilon_\eta=0$, we find that
the distorted $\nue$ spectrum is very accurately reproduced by the equal
fluxes case with a temperature 
\begin{equation} 
T_x \approx (1-f_L) T_{\mu} + (1-P_H) f_L T_{\tau}
\label{tx}
\end{equation} 
of the $\nux$ spectrum. In eq. (\ref{tx}) the $\numu$ and $\nutau$
temperatures are weighted by the same coefficients that multiply the
corresponding fluxes in the expression of the observed $\nue$ flux,
eq.   (\ref{explicit}). 
If $\epsilon_T>0$ ($\epsilon_T<0$), we have $T_\mu < T_x < T_\tau $
($T_\tau < T_x < T_\mu$) and the differences  $| T_x - T_\mu|$ and  $| T_x
- T_\tau|$ are well within the uncertainty of the predicted values of
$T_\mu$ and $T_\tau$. 
A numerical example is given in fig. \ref{fig:f6}, which shows the energy
spectrum of events at the SNO detector from the  $\nue + d \rightarrow e^-
+ p + p$ detection process 
\footnote{Events from the $\barnue + d \rightarrow e^+ + n + n$
reaction can in principle be distinguished from those from $\nue + d
\rightarrow e^- + p + p$ due to the detection of at least one neutron
by capture on deuterium, with characteristic time $t_{n d}\sim 5$ ms,
in coincidence with the charged lepton.  The coincidence can be
established with high efficiency ($\sim 75-80\%$ \cite{Steiger:cm,JK})
due to the relatively small event rates of the two processes at SNO
($\sim 100-200$ events for a supernova at $D=10$ kpc distance, see
e.g. \cite{Takahashi:2001ep}), which guarantees the average time
interval between two events to be significantly larger than $t_{n
d}$.}  
with equal and different $\numu$-$\nutau$ fluxes. We considered
thermal time-integrated spectra as in eq. (\ref{eq3}) with no
pinching, equal integrated luminosities $E_\mu=E_\tau=5\cdot
10^{52}~{\rm ergs}$ and a distance $D=8.5$ kpc from the supernova.
LMA oscillation parameters ($\tan^2 \theta_{12}=0.33$) were assumed,
together with normal mass hierarchy and $P_H=0$. As follows from the
figure, the $\nue$ spectrum obtained with different temperatures,
$T_\mu=7$ MeV and $T_\tau=8.4$ MeV, is well reproduced by that with
the common temperature $T_x=7.3$ MeV, according to eq. (\ref{tx}).

Global fits of simulated SNO and SuperKamiokande data \cite{Barger:2001yx}, 
performed under the assumption of equal temperatures of all the non-electron
neutrino species and exact equipartition of the total energy $E_B$,  allow 
to determine these parameters with precision better than $\sim 10\%$. 
Considering that the data set is dominated by the high statistics in
SuperKamiokande, this result should be interpreted as a determination of
the effective  temperature and integrated luminosity of the non-electron
antineutrinos, $T_x$ and $E_x$. 

It can be checked that, if the $\numu$ and $\nutau$ spectra have different
pinching factors and luminosity difference of $\sim 10 -20 \%$, the equal
fluxes case  accounts very well for the observed $\nue$ spectrum once 
effective values of $T_x$ and  $E_x$  are chosen within their uncertainty
intervals, as they could be measured by SuperKamiokande.

In principle, the possibility exists that the differences between the 
$\numu$ and $\nutau$ (or $\barnumu$ and $\barnutau$) fluxes 
will be probed at future large detectors (UNO \cite{Jung:1999jq},
HyperKamiokande \cite{HyperK}, etc.) once all the neutrino oscillation
parameters are known to a sufficient accuracy.

%%%%%%%%%%%%%%%%%%%%%%%%%%%%%%%%%%%%%%%%%%%%%%%%%%%%%%%%%%%%%%%%%%%%%%%%%%%
%%%%%%%%%%%%%%%%%%%%%%%%%%%%%%%%%%%%%%%%%%%%%%%%%%%%%%%%%%%%%%%%%%%%%%%%%%%
\section{Earth matter effects}
\label{sect:4}
%%%%%%%%%%%%%%%%%%%%%%%%%%%%%%%%%%%%%%%%%%%%%%%%%%%%%%%%%%%%%%%%%%%%%%%%%%%
%
%
%%%%%%%%%%%%%%%%%%%%%%%%%%%%%%%%%%%%%%%%%%%%%%%%%%%%%%%%%%%%%%%%%%%%%%%%%%%
\subsection{Unequal fluxes and conversion in the Earth}
\label{sect:4.1}

If the neutrino burst crosses the Earth before detection, the fluxes of
the various neutrino flavors in the detector are affected by the
regeneration in the matter of the Earth.

Let us first consider these effects in the
scheme with normal mass hierarchy. 
According to a simple generalization of eqs. (\ref{final}) and
(\ref{Pmatrix}), the fluxes  ${\vec F^D}\equiv \left(F^D_e,F^D_{\mu},
F^D_\tau \right)$ of the neutrino flavor states in the detector can be
expressed as
\begin{equation}
{\vec F^D}=Q^\oplus {\vec {F^0 }}~,\quad\quad\quad
%\label{finalearth}  \\
Q^\oplus\equiv S^\oplus {\cal P}~,
\label{Pmatrixearth}
\end{equation} 
where the matrix ${\cal P}$ was given in eq. (\ref{calp}). Each entry
$S^\oplus_{\alpha i}$ of the matrix $S^\oplus$ represents the probability
that a neutrino arriving at the Earth as $\nu_i$ interacts, 
after crossing the Earth,  as a $\nu_\alpha$
in the detector: $S^\oplus_{\alpha i}\equiv P(\nu_i \rightarrow
\nu_\alpha)$.  From eqs. (\ref{calp}) and (\ref{Pmatrixearth}) 
the  observed $\nue$ flux equals: 
\begin{equation}
F^D_{ e} = F_{ e}^0 P_H  f^\oplus_L + F_{ \mu}^0 (1-f^\oplus_L) + F_{
\tau}^0 (1-P_H) f^\oplus_L~,
\label{explE}
\end{equation}
where
\begin{equation}
f^\oplus_L\equiv P_L + (1-2 P_L) S^\oplus_{e 2}~.
\label{floplus}
\end{equation}
In eqs. (\ref{explE})-(\ref{floplus}) terms of order $\sin^2 \theta_{13}$
were neglected \footnote{It can be checked \cite{Dighe:1999bi} that the
contributions of these terms to the difference (\ref{eartheff}) are
smaller than $10^{-3}$, and therefore negligible.  }.

Combining eqs. (\ref{fl}), (\ref{explicit}) and (\ref{explE})-(\ref{floplus}) 
we obtain the deviation due to the Earth matter effects of the observed
$\nue$ flux, $F^D_{ e}$, from that predicted for conversion in the star
only: 
\begin{equation}
F^D_{ e}-F_{ e}=(1-2P_L) f_{reg} \left[ P_H F_{ e}^0 - F_{ \mu}^0+
(1-P_H) F_{ \tau}^0 \right]~, 
\label{eartheff}
\end{equation}
where the {\it regeneration factor},
\begin{equation}
f_{reg}\equiv S^\oplus_{e 2}-\sin^2 \theta_{12}~,
\label{regf}
\end{equation}
accounts for the matter effects inside the Earth. It depends on the
solar neutrino parameters, the neutrino energy, 
and the direction of the neutrino trajectory inside
the Earth. 

In what follows we consider $\Delta m^2_{21}$ and $\theta_{12}$ in the
LMA region.  (The Earth matter effect 
is small for the other solutions of the solar neutrino problem; see
e.g. the discussion in ref. \cite{Lunardini:2001pb}). 
For the LMA oscillation parameters the conversion in the L
resonance is adiabatic, thus we set $P_L=0$ from now on.

As can be seen from eq. (\ref{eartheff}), the contribution of $F_{\tau}^0$
to the difference $F^D_{ e}-F_{ e}$ is small for large violation of
adiabaticity in the H resonance and vanishes for $P_H=1$, according to the
discussion in sec. \ref{sect:3.2}. In this case the difference of
the $\numu$ and $\nutau$ fluxes clearly has no observable effects.  
Conversely, in the case of good adiabaticity, $P_H=0$, the contribution
of the $\nue$ flux, $F_{ e}^0$, is absent and the whole Earth effect is
due to the difference $F_{ \tau}^0-F_{ \mu}^0$. 

It is useful to split the expression (\ref{eartheff}) into two
terms according to their dependence on $P_H$:  
\begin{equation}
F^D_{ e}-F_{ e}=\Delta F^{(2)}_e +\Delta F^{corr}_e~,  
\label{sep} 
\end{equation}
where
\begin{equation}
\Delta F^{(2)}_e =  f_{reg} P_H \left( F_{ e}^0-F_{ \tau}^0 \right)  ~,       
%\label{main}  \\
\quad\quad
\Delta F^{corr}_e =  f_{reg}  (F_{ \tau}^0-F_{ \mu}^0)~. 
\label{corr}
\end{equation}

\noindent
Let us comment on the quantities $\Delta F^{(2)}_e$ and $\Delta F^{corr}_e$.

\begin{itemize}

\item $\Delta F^{(2)}_e$ describes the whole Earth matter effects if the
$\numu$--$\nutau$ fluxes are equal. 
Being proportional to the difference $F_{ e}^0-F_{ \tau}^0$, it changes
the sign at the critical energy $E^{(2)}_C$ discussed in
sec. \ref{sect:3.3}. 
If  the regeneration factor $f_{reg}$ is positive (as it is for conversion
in the mantle of the Earth  only \cite{Dighe:1999bi,Lunardini:2001pb}) we 
have  $\Delta F^{(2)}_e>0$
for $E<E^{(2)}_C$ and $\Delta F^{(2)}_e<0$ for $E>E^{(2)}_C$.
 The quantity $\Delta F^{(2)}_e$ is also proportional to $P_H$ and
vanishes if $P_H=0$, according to the fact that if the H resonance is
completely adiabatic the flux $F^0_e$ does not undergo the L
resonance.  Conversely,  $\Delta F^{(2)}_e$ is maximal if  $P_H=1$.

\item $\Delta F^{corr}_e$ is proportional to the
difference of the $\numu$ and $\nutau$ fluxes.  
%therefore it is a  measure
%of the $\numu$--$\nutau$ inequivalence. 
As discussed in sec. \ref{sect:2.2},  $\Delta F^{corr}_e$ changes sign at
$E= E^{corr}_C$. If  $f_{reg}$ is positive and $\epsilon_T>0$ we have
$\Delta F^{corr}_e<0$ for $E<E^{corr}_C$ and $\Delta F^{corr}_e>0$  for
$E>E^{corr}_C$. The opposite signs are realized for $\epsilon_T<0$, while
there is no change  of sign  (see sec. \ref{sect:2.2}) if the $\numu$ and
$\nutau$ fluxes differ only in the luminosity. 

\end{itemize}

The relative size of $\Delta F^{(2)}_e$ and $\Delta F^{corr}_e$ depends on
$P_H$. In particular, in the high energy limit, $E\gg E^{(2)}_C$, where
$F^0_e/F^0_\mu \ll 1$ and the regeneration factor is large  
\cite{Lunardini:2001pb}, the condition $|\Delta F^{corr}_e/\Delta
F^{(2)}_e|\gta 1$ requires 
\begin{eqnarray} 
&&P_H \lta P^{eq}_H 
\label{cond} \\ 
&&P^{eq}_H\equiv \left|  \frac{ \Delta_F}{ 1+ \Delta_F  }  \right|~,
\label{peq}
\end{eqnarray}
as follows from eqs. (\ref{corr}) and (\ref{reldiff}). Taking  
$E=60$ MeV, $T_{\mu}=7$ MeV $\epsilon_L=\epsilon_\eta=0$  and
$\epsilon_T=0.1$  one  gets $P^{eq}_H\simeq 0.33$. For $P_H \gta P^{eq}_H$
(i.e.  $P_H \sim 1$) the Earth matter effect is dominated by the $\Delta
F^{(2)}_e$ term, while if $P_H \ll P^{eq}_H$ the larger contribution to
the effect is given by $\Delta F^{corr}$, which, as we have discussed, is
entirely due to the differences in the $\numu$--$\nutau$ fluxes. 

%Similarly to the results of sec. \ref{sect:3.3}, we found that 
For $P_H \gta P^{eq}_H$, the effect of the $\numu$ -- $\nutau$ flux  
difference 
is very difficult to identify due to the uncertainties in the values of
the various parameters: for $|\epsilon_T|,|\epsilon_T|\lta 0.2$ the flux
$F^D_e$ observed in the detector is very accurately reproduced by the
equal $\numu$--$\nutau$ fluxes case with values of the parameters 
%(e.g. the matter density profile of the star, which enters in the
%determination of the value of $P_H$) 
within their uncertainties.
In contrast to this, if $P_H \ll P^{eq}_H$ the term $\Delta F^{corr}_e$ 
due to the unequal fluxes dominates, and in the limit $P_H=0$ it 
represents the only Earth matter effect. The absence of competing effects 
offers the possibility of probing the differences in the $\numu$ -- $\nutau$
fluxes and therefore makes this case of particular interest. 
This is true especially if $P_H$ is known to be very small (i.e. $\sin^2
\theta_{13}\gta 10^{-3}$) from independent measurements provided, e.g., by 
neutrino factories or superbeams \cite{Albright:2000xi}.  
A possibility is also related to  the comparison of charged current and
neutral current events from the neutronization burst at SNO
\cite{Barger:2001yx}.  If the value of $P_H$ is unknown, the possibility
that the Earth effect is due to $F^0_{\mu} \neq F^0_{\tau}$ with $P_H=0$
could be distinguished, in principle, from that of equal fluxes with a
non-zero $P_H$ (eq. (\ref{corr})) at least when the 
 two effects have different signs at high energies, i.e. for
$\epsilon_T>0$ and/or $\epsilon_L>0$.
\\

The description of the Earth matter effects on the conversion of antineutrinos 
with inverted mass hierarchy is analogous to that presented above.  
Similarly
to eqs. (\ref{sep})-(\ref{corr}), the difference of the 
observed $\barnue$ fluxes with and without Earth effects is given by 
\begin{equation}
F^D_{\bar e}-F_{\bar e}=\Delta F^{(2)}_{\bar e} +\Delta F^{corr}_{\bar e}~,  
\label{sepb} 
\end{equation}
\begin{equation}
\Delta F^{(2)}_{\bar e}= {\bar f}_{reg} P_H \left( F_{\bar e}^0-F_{\bar
\tau}^0 \right)  ~,         
%\label{mainb}  \\
\quad\quad
 \Delta F^{corr}_{\bar e} =  {\bar f}_{reg}  (F_{\bar \tau}^0-F_{\bar
\mu}^0)~, 
\label{corrb}
\end{equation}
where
\begin{equation}
\bar f_{reg} \equiv S^\oplus_{\bar e 1}-\cos^2 \theta_{12}~,
\label{regfanti}
\end{equation}
and  $S^\oplus_{\bar e 1}$ is the $\bar \nu_1 \rightarrow \barnue$
conversion probability in the matter of the Earth.
The results obtained above for neutrinos can be immediately extended to
antineutrinos.  However, compared to the neutrino channel, the regeneration 
factor $\bar f_{reg}$ is smaller in the antineutrino  case, thus giving a 
smaller  effect (see e.g. \cite{Lunardini:2001pb}). 

It should be noted that, although the potential $V_{\tau\mu}$ is induced
by radiative corrections and  therefore is much smaller than the potential
$V_e$, it plays a very important role in the evolution of neutrino states in
the star provided that the original fluxes of $\numu$ and $\nutau$ are
different. If one had disregarded it, one would have had an additional 
suppression factor $\cos 2\theta_{23}\ll 1$ in the expressions for 
$\Delta F_e^{corr}$ and $\Delta F_{\bar{e}}^{corr}$ and so much smaller 
Earth matter effects.

%%%%%%%%%%%%%%%%%%%%%%%%%%%%%%%%%%%%%%%%%%%%%%%%%%%%%%%%%%%%%%%%%%%%%%%%%%%
\subsection{Observable signals}
\label{sect:4.2}

\begin{figure}[hbt]
\begin{center}
\epsfig{file=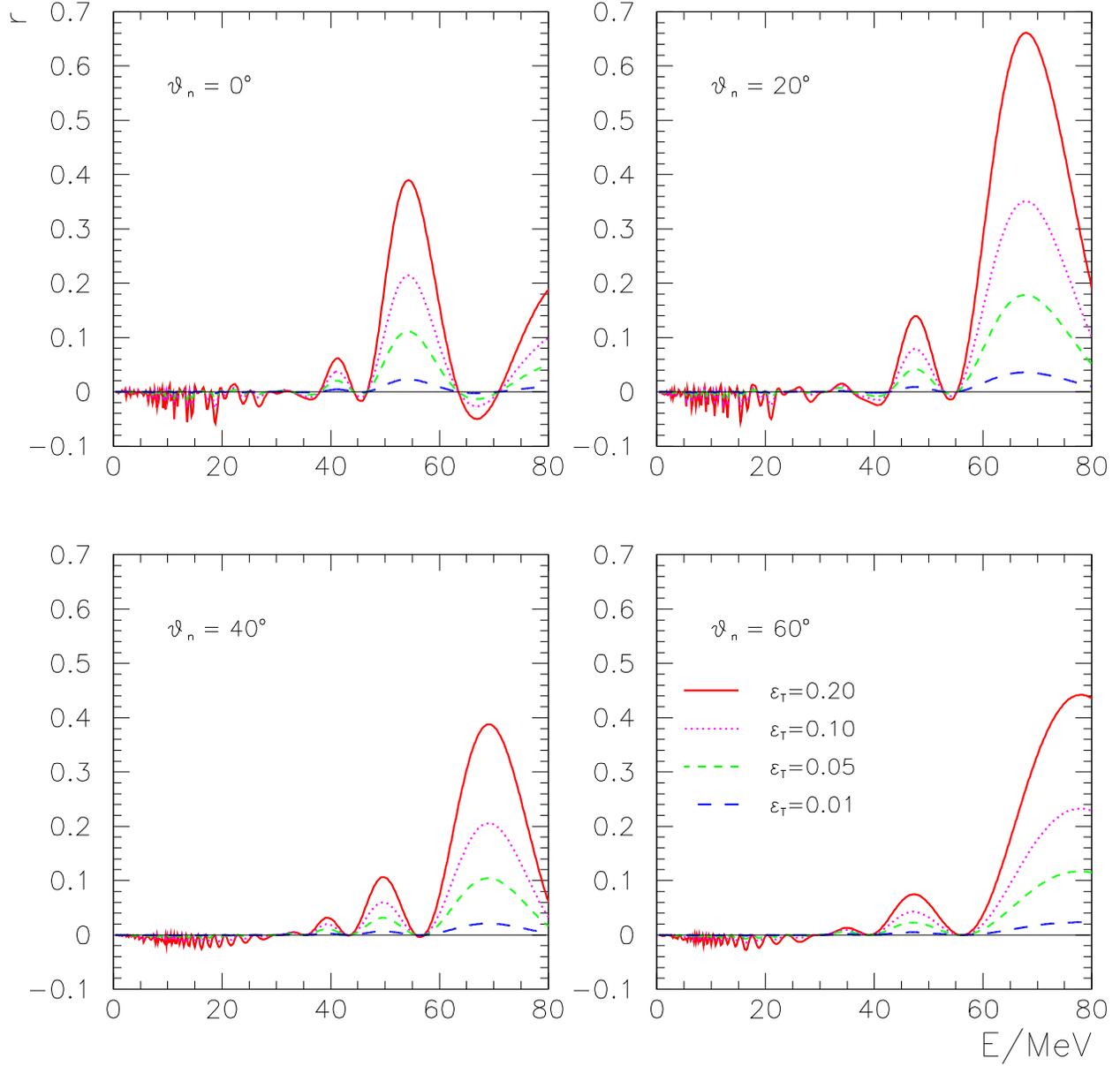, width=18truecm}
\end{center}
\caption{The relative Earth matter effect in the neutrino channel, $r$, as
a function of the neutrino energy for different values of the trajectory
nadir angle, $\theta_n$, and for a number of values of the relative
$\numu$ -- $\nutau$ temperature difference $\epsilon_T$. We considered
normal mass hierarchy, $P_H=0$, $\epsilon_L=0$, $T_\mu=7$ MeV, $\sin^2
2\theta_{12}=0.75$ and $\Delta m^2_{21}=5\cdot 10^{-5}~{\rm eV^2}$.}
\label{fig:f1}
\end{figure}

\begin{figure}[hbt]
\begin{center}
\epsfig{file=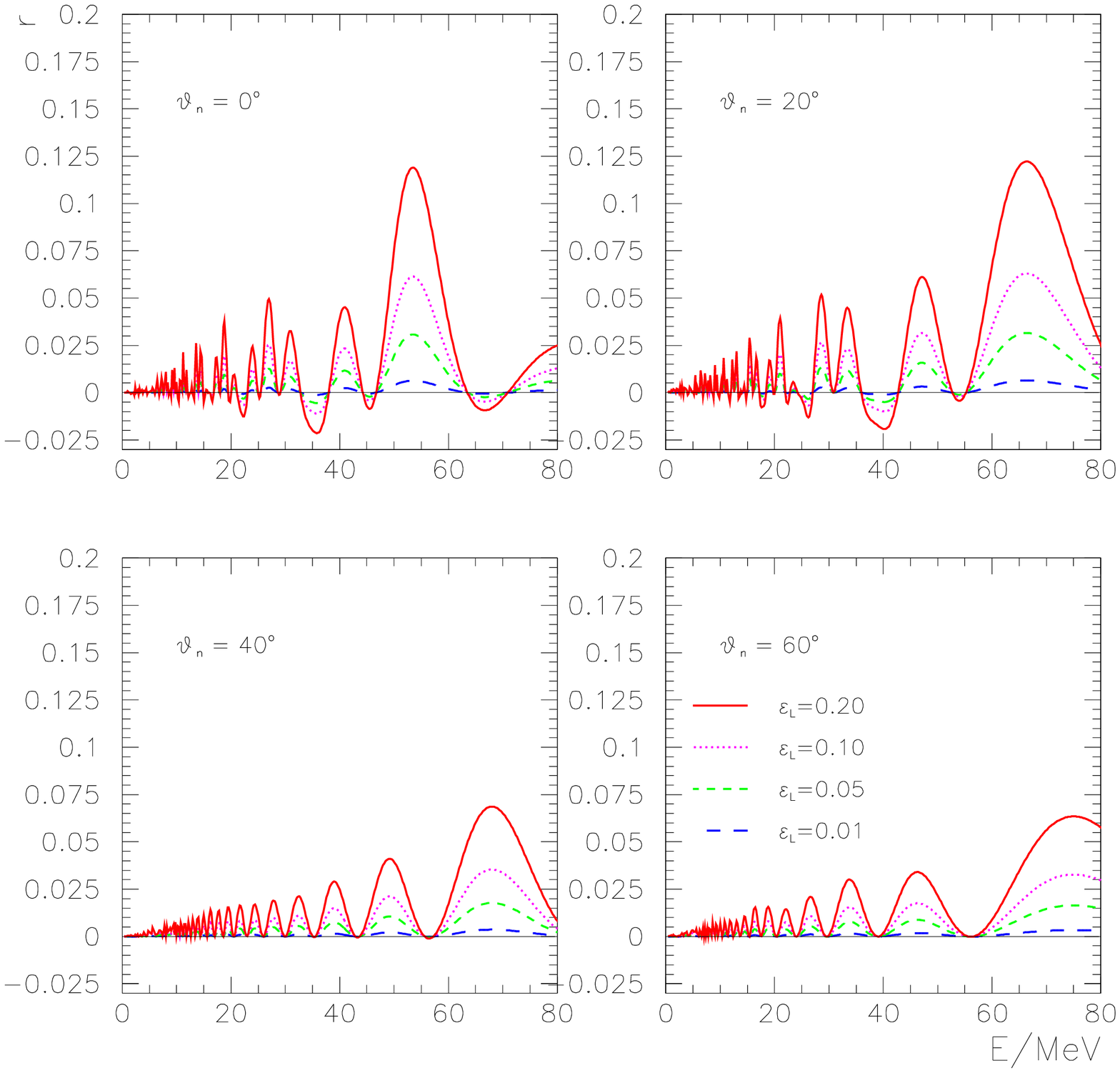, width=18truecm}
\end{center}
\caption{The same as in fig. \ref{fig:f1} but for $\epsilon_T=0$ and
different values of the relative differences of  luminosities, $\epsilon_L$.}
\label{fig:f8}
\end{figure}

\begin{figure}[hbt]
\begin{center}
\epsfig{file=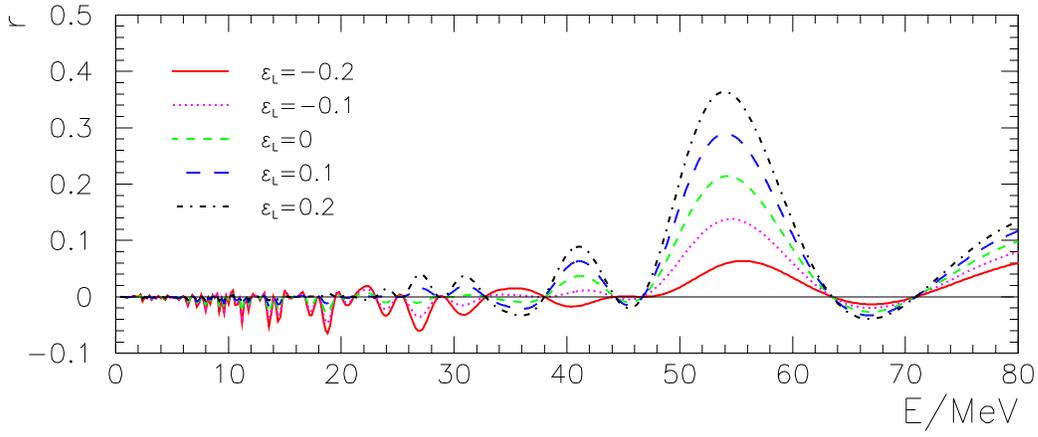, width=15truecm}
\end{center}
\caption{The relative Earth matter effect in the neutrino channel, $r$, as
a function of the neutrino energy for  $\epsilon_T=0.1$ and different
values of $\epsilon_L$. We have taken  $\theta_n=0^\circ$; the values of
the other parameters are the same as in fig. \ref{fig:f1}.}
\label{fig:f9}
\end{figure}

\begin{figure}[hbt]
\begin{center}
\epsfig{file=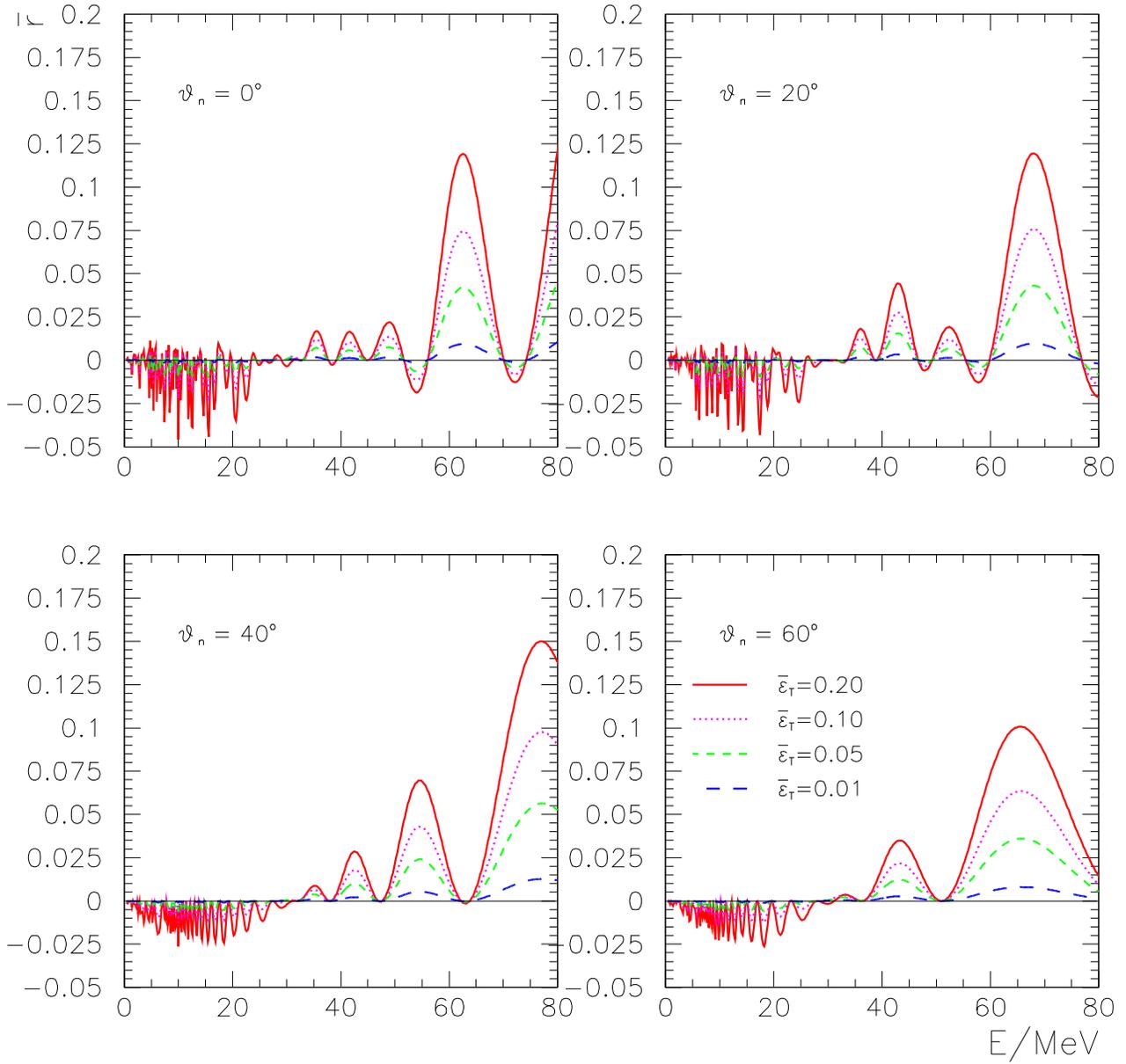, width=18truecm}
\end{center}
\caption{The same as in fig. \ref{fig:f1} but for the antineutrino channel
and inverted mass hierarchy. }
\label{fig:f2}
\end{figure}

\begin{figure}[hbt]
\begin{center}
\epsfig{file=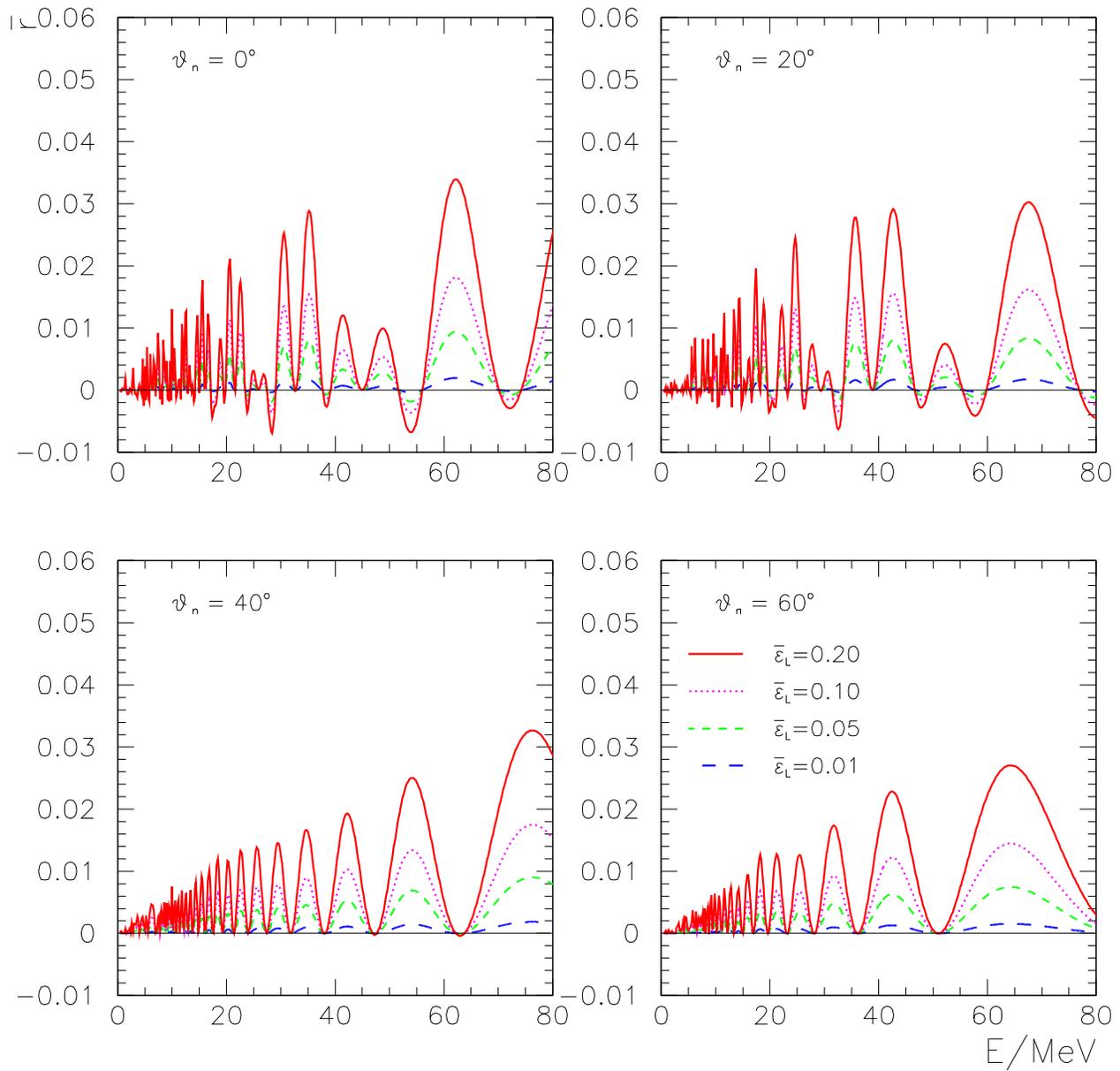, width=18truecm}
\end{center}
\caption{The same as in fig. \ref{fig:f2} but for ${\bar \epsilon}_T=0$
and different values of the relative differences of the integrated
luminosities ${\bar \epsilon}_L$.}
\label{fig:f3}
\end{figure}

\begin{figure}[hbt]
\begin{center}
\epsfig{file=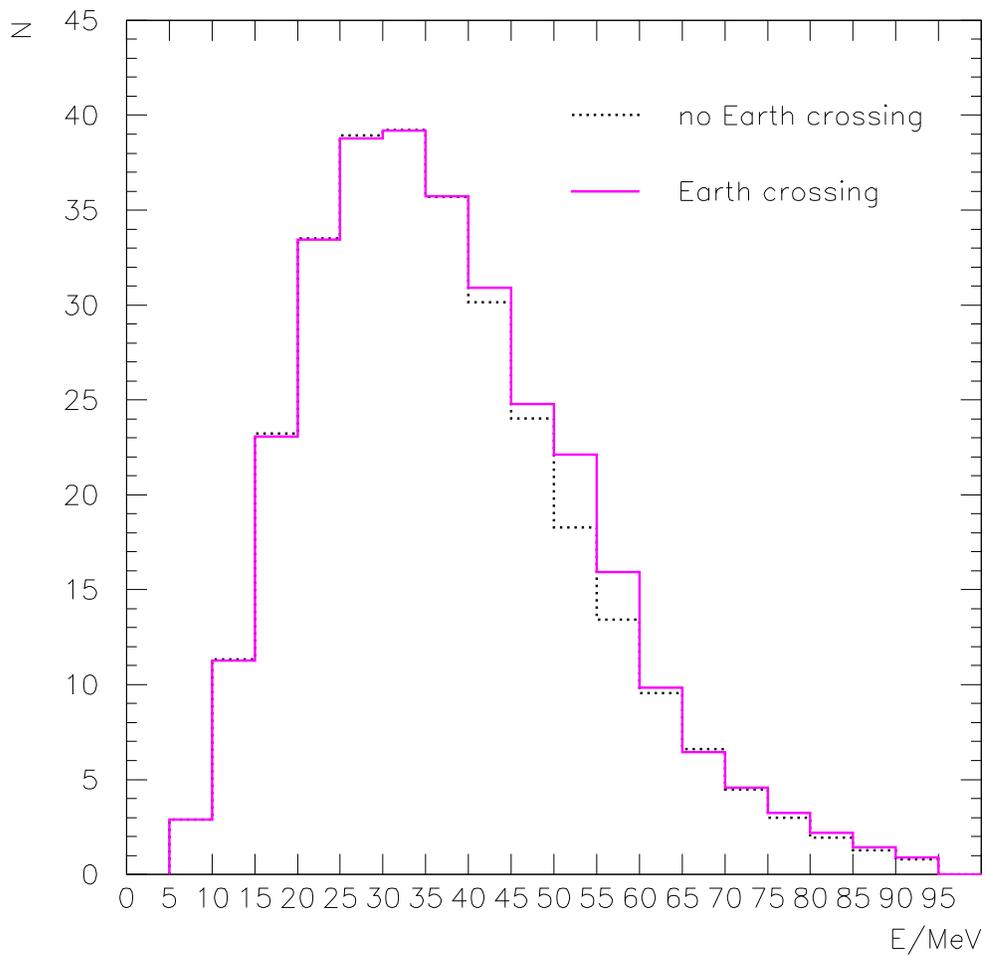, width=15truecm}
\end{center}
\caption{The spectrum of events expected at SNO from the $\nue + d
\rightarrow e^- + p + p$ process with and without Earth crossing. We took
$\epsilon_L=0$ and $\epsilon_T=0.1$, a distance to the supernova $D=8.5$
kpc, nadir angle $\theta_n=0^\circ$ and integrated luminosities
$E_\mu=E_\tau=5\cdot 10^{52}~{\rm ergs}$. The  other parameters are the
same as in fig. \ref{fig:f1}.}
\label{fig:f4}
\end{figure}

\begin{figure}[hbt]
\begin{center}
\epsfig{file=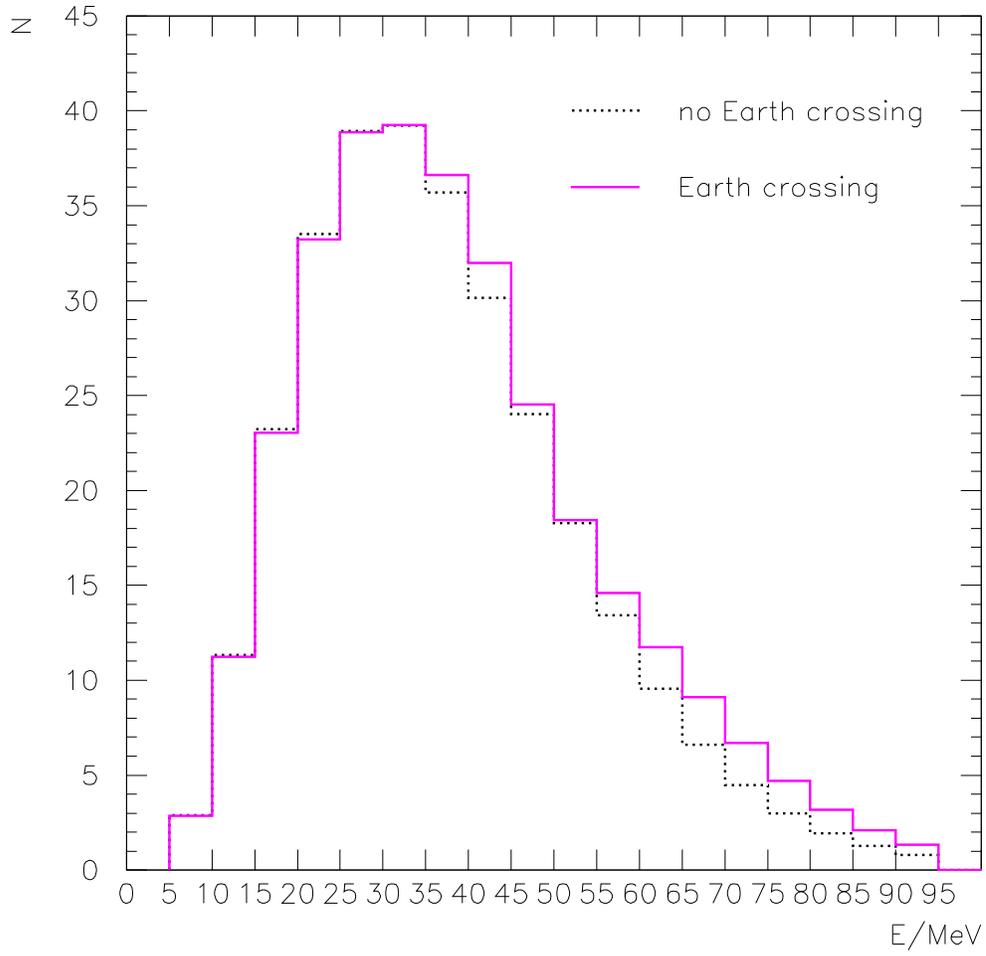, width=15truecm}
\end{center}
\caption{The same as fig. \ref{fig:f4} but for $\Delta m^2_{21}=3\cdot
10^{-5}~{\rm eV^2}$ and $\theta_n=40^\circ$.}
\label{fig:f5}
\end{figure}

Let us now discuss the sensitivity of the Earth matter effect to the
$\numu$--$\nutau$ flux differences. 
%In the light of  the results of sec. \ref{sect:4.1}, 
Consider the effect in the neutrino channel for normal mass hierarchy,  
solar neutrino parameters in the LMA  region ($P_L = 0$) and $\sin^2
\theta_{13}\gta 10^{-3}$ (i.e. $P_H\simeq 0$). In this case the relative
Earth matter effect, $r\equiv (F^D_e-F_e)/F_e$, is entirely due to the 
differences in the $\numu$--$\nutau$ fluxes:
\begin{equation}
r= \frac{ \Delta F^{corr}_e}{ F_e  }~,
\label{releff0}
\end{equation}
and, from eqs. (\ref{reldiff}), (\ref{corr}) and (\ref{decflux}) one gets 
\begin{equation}
r =  \frac{ \Delta_F}{ 1+   s^2_{12} \Delta_F  }  f_{reg}~.
\label{thatis} 
\end{equation}
A study of the ratio $r$ is presented in figs. \ref{fig:f1}-\ref{fig:f5}, 
in which the cases of different $\numu$ and $\nutau$ temperatures and
luminosities are considered. The effect of different pinching factors, being 
energy-independent at high energy, $E\gg E^{(2)}_C$ (see sec. \ref{sect:2.1}), 
is equivalent to a difference in the luminosities and therefore will
not be discussed here.   \\

Let us first consider the situation in which the 
difference between the $\numu$ and $\nutau$ fluxes is mainly due to
different temperatures and take $\epsilon_T>0$ and $\epsilon_L=0$.  
The ratio $r$ depends on the neutrino energy via the regeneration factor
$f_{reg}$ and via the difference $\Delta_F$, eqs. (\ref{reldiff}) and
(\ref{expa}).  The regeneration factor determines the oscillatory behavior
of $r$ with the energy, while the dependence of $r$ on $\Delta_F$
leads to the change of the sign of $r$ at $E=E^{corr}_C$. This sign  
changes from negative at lower energies to positive at higher energies,
provided that $f_{reg}>0$. Both  $f_{reg}$ and $\Delta_F$ increase with 
energy (see eq. (\ref{expa})), therefore the largest effect is achieved in
the high energy part of the spectrum. It also increases with increasing  
$\epsilon_T$. These features are illustrated in fig. \ref{fig:f1}, 
which shows $r$ as a function of the energy $E$ for 
different values of $\epsilon_T$ and of the nadir angle $\theta_n$  of the
neutrino trajectory in the Earth.  The figure has been obtained using a
realistic density profile of the Earth \cite{PREM}. 
%The parameters
%$\sin^2 2\theta_{12}=0.75$, $\Delta m^2_{21}=5 \cdot 10^{-5}~ {\rm eV}^2$
%and $T_\mu=7$ MeV have been taken. 
As fig. \ref{fig:f1} shows, the oscillatory distortions of the neutrino
energy spectrum have larger periods for larger $E$, due to the larger
oscillation length in matter, and for larger  nadir angle $\theta_n$ 
(i.e. shorter trajectory inside the Earth). The size of the effect increases
with the energy due to the linear increase of $\Delta_F$ with $E$ (see
eq. (\ref{expalim})) and to the larger regeneration factor. 
Taking into account that $f_{reg}\lta 0.7$ \cite{Lunardini:2001pb} and
that $r/f_{reg}$ can be as large as $r/f_{reg}\sim 0.45$ for $E\sim 60 -70$ MeV
and $\epsilon_T\sim 0.1$, we find $r \lta 0.3$.
%as it appears in the figure.  
For larger temperature difference, $\epsilon_T\sim 0.2$, the effect can
reach $\sim 70\%$. 

If $\epsilon_T<0$, $r$ is positive for $E < E^{corr}_C$ and
negative for $E > E^{corr}_C$. Besides this change of the
sign, the general features and the size of the effect are  
similar to those discussed above for  positive $\epsilon_T$. 
This is expected, considering that $r$ is linear in $\epsilon_T$ when the
latter is small (see eqs. (\ref{expa}) and (\ref{thatis})). 
The linear approximation is good for $|\epsilon_T|\lta 0.05 - 0.1$. 
\\

Let us now consider effect of different luminosities: $\epsilon_T=0$ 
and $\epsilon_L\neq 0$. In this case we have $\Delta_F=\epsilon_L$ 
(eq. (\ref{expa})), and therefore 
$r\simeq \epsilon_L f_{reg}$, as follows from eq. (\ref{thatis}). The only
energy dependence comes from the regeneration factor $f_{reg} $. For the 
luminosity differences $|\epsilon_L|\lta 0.2$  and $f_{reg} \simeq 0.7$ we
find $|r|\lta 0.15$. This is confirmed by the results shown in the fig. 
\ref{fig:f8}.
\\

In general, both the temperatures and the integrated luminosities of the
$\numu$ and $\nutau$ fluxes may differ.  The interplay of these two
differences can enhance or suppress the Earth matter effect $r$ compared 
to the cases we have discussed above.   Since, as we have shown, the larger
contribution to the Earth matter effect comes from the temperature
difference, it is illuminating to consider the effect of a small
luminosity difference for a given difference $\epsilon_T$ of the
temperatures. We focus on the high-energy part of the spectrum because 
at low energies the regeneration factor $f_{reg}$ is small.

As can be understood from eqs. (\ref{rat}), (\ref{expa}) and
(\ref{thatis}), at high energies the relative difference of fluxes
$\Delta_F$, and therefore the ratio $r$, is larger (in absolute
value) when $\epsilon_L$ and $\epsilon_T$ have the same
sign.

Conversely, if $\epsilon_L$ and $\epsilon_T$ have opposite signs, the
effects of the two differences in eq. (\ref{expa}) tend to cancel each other. 
%This partial compensation, which is energy dependent, reduces the 
%relative Earth matter effect $r$. 
This is illustrated in 
fig. \ref{fig:f9}, which shows the ratio $r$ as a function of energy
for $\epsilon_T=0.1$ and different values of $\epsilon_L$, with the same
parameters as in fig. \ref{fig:f1}.  From the figures it follows that the
enhancement or suppression of the Earth matter effect due to the 
$\numu$ -- $\nutau$ luminosity difference can be as large as $40 - 50\%$, 
depending on the neutrino energy.
\\

If the mass hierarchy
is inverted and $P_H=0$, the Earth matter effect due to the $\barnumu$ --
$\barnutau$ flux difference, $\Delta F^{corr}_{\bar e}$, appears in the
antineutrino channel without competing effects 
($\Delta F^{(2)}_{\bar e}=0$, see  sec. \ref{sect:4.1}).
The expression for the relative Earth matter effect $\bar r$ can be
obtained from eqs. (\ref{finalb}), (\ref{xQ0}) 
and (\ref{sepb})-(\ref{corrb}):
\begin{equation}
{\bar r} =  \frac{{\bar \Delta}_F}{ 1+  c^2_{12} {\bar \Delta}_F  }
{\bar f}_{reg} ~.  
\label{thatis2}
\end{equation}  
As can be seen from eqs. (\ref{thatis}) and (\ref{thatis2}), for equal
relative differences of fluxes, $\Delta_F={\bar \Delta}_F$, the
quantity $\bar r$ differs from $r$ by small terms of order ${\bar
\Delta}^2_F$ and by the suppression factor $\bar f_{reg} /f_{reg}<1$.
For the antineutrino channel plots analogous to those in
figs. \ref{fig:f1} and \ref{fig:f8} are presented in figs.
\ref{fig:f2} and \ref{fig:f3}.  \\

As a further illustration, in figs. \ref{fig:f4} and \ref{fig:f5}  we show
the binned energy spectra of the events expected in the heavy water tank
of  SNO for different $\numu$ and $\nutau$ fluxes with and without Earth
crossing. 
In the calculation of the  expected numbers of events the detection
efficiency and energy resolution have been taken into account following
the discussion in ref. \cite{Lunardini:2001pb}. In fig. \ref{fig:f4} we 
considered unpinched spectra  with $\epsilon_L=0$, $\epsilon_T=0.1$ and
the other neutrino parameters the same as in
fig. \ref{fig:f1}. The nadir angle $\theta_n=0^\circ$ was taken; the
distance $D$ and the integrated luminosities are the same as in
fig.  \ref{fig:f6}. Fig. \ref{fig:f5} shows the same as fig. \ref{fig:f4}
but with $\Delta m^2_{21}=3\cdot 10^{-5}~{\rm eV^2}$ and $\theta_n=40^\circ$. 

As follows from the figures, the Earth matter effect caused by the
difference of $\numu$ and $\nutau$ fluxes leads to a distortion of the
observed energy spectrum as a whole or in isolated energy bins,
depending on the depth of the neutrino trajectory in the Earth. This
distortion amounts to $10 - 20 \%$ at most, and, due to the expected
limited number of events, is not statistically significant.  This
deviation could have the significance of $3\sigma$ or larger for a
supernova at a smaller distance (2-3 kpc) and/or in a larger
experiment. For instance the proposed liquid argon detector LANNDD is
expected to observe about 3000 $\nue$ events \cite{Cline:2001pt} --
more than 10 times larger statistics than SNO -- for the
parameters used in figs. \ref{fig:f4} and \ref{fig:f5}.  The
sensitivity of this detector to the $\nue$ energy spectrum, however,
remains to be investigated. Moreover, as discussed in
\cite{Lunardini:2001pb}, establishing the Earth matter effects may
require the comparison of the spectra observed by the detectors at
different locations, and at present it is not clear if more than one
large $\nue$ detector with the necessary characteristics\footnote{Besides 
the high statistics, the observation of the Earth matter effect requires
\cite{Lunardini:2001pb}: (i) separate detection 
of $\nue$ and $\nue+\numu+\nutau$ signals, (ii) separate detection of
neutrinos and antineutrinos, and (iii) the reconstruction of the
energy spectrum.} will be built in future.
\\

The Earth matter effect in the antineutrino channel, eq. (\ref{thatis2}),
can be detected via the $\barnue + p \rightarrow e^+ + n$ reaction in
Cherenkov or scintillator detectors 
\footnote{We checked that the contributions of charged current
scatterings on nuclei, e.g. $\nue + ^{16}O \rightarrow ^{16}F + e^-$,
to the event rates is smaller than $2-4\%$ and therefore negligible.}.
The distortions of the observed energy spectrum compared to the case
of no Earth crossing by neutrinos do not exceed $10\%$. Establishing
this difference with at least $3 \sigma$ statistical significance
would require a comparison of data from two large detectors. We find
that, unless the distance to the supernova is very small ($D=$2 -- 3
kpc), two detectors of SuperKamiokande (SK) size do not have enough
sensitivity. In contrast to this, large water Cherenkov detectors of
next generation like UNO \cite{Jung:1999jq}, HyperKamiokande
\cite{HyperK} or TITAND \cite{Suzuki:2001rb} ($\sim 18$, $ 40$ and
$60$ times larger than SK respectively) could establish the effect
with significance $3 \sigma$ or larger for distances $D$ up to $\sim
10 $ kpc at least.  \\

The results for the time-integrated energy spectra presented here can be
extended to the spectra of events collected in any (small) time interval
within the duration of the burst. Taking into account that the difference
between the fluxes of the original non-electron neutrinos may vary with
time, a study of the time dependence of the signal would be of interest to
test the effect we have discussed.  This is especially true if a specific 
time dependence of the $\numu$ -- $\nutau$ flux difference is predicted by
supernova models.

%%%%%%%%%%%%%%%%%%%%%%%%%%%%%%%%%%%%%%%%%%%%%%%%%%%%%%%%%%%%%%%%%%%%%%%%%%%
%%%%%%%%%%%%%%%%%%%%%%%%%%%%%%%%%%%%%%%%%%%%%%%%%%%%%%%%%%%%%%%%%%%%%%%%%%%
\section{Discussion and conclusions}
\label{sect:5}
%%%%%%%%%%%%%%%%%%%%%%%%%%%%%%%%%%%%%%%%%%%%%%%%%%%%%%%%%%%%%%%%%%%%%%%%%%%

If the solution of the solar neutrino problem is in the LMA region,
the Earth matter effects appear in the spectra of $\nue$ or $\barnue$ from 
a supernova or in both, depending on the type of the neutrino mass
hierarchy, mixing angle $\theta_{13}$ and possible differences between 
the $\numu$, $\nutau$, $\barnumu$ and $\barnutau$ fluxes.  In
particular, if (i) $F^0_\mu=F^0_\tau $ ($F^0_{\bar \mu}=F^0_{\bar \tau}$),
(ii) the hierarchy is normal (inverted), and (iii) the conversion in the H
resonance is adiabatic, $P_H\simeq 0$ (i.e. $\sin^2 \theta_{13}>
10^{-3}$), the Earth matter effects exist in the $\barnue$ ($\nue$) channel 
only \cite{Dighe:1999bi,Lunardini:2001pb}.

This is no longer true if there are differences between the original
$\numu$ and $\nutau$ or $\barnumu$ and $\barnutau$ fluxes. If $F^0_\mu\neq 
F^0_\tau $ ($F^0_{\bar \mu}\neq F^0_{\bar \tau}$) we find that, even for
$P_H=0$, effects of regeneration in the  matter of the Earth as large as 
$\sim 70  \%$ ($\sim 20 \%$) can take place in the $\nue$ ($\barnue$) 
spectrum for normal (inverted) hierarchy.

Therefore, if $P_H$ is known to be very small from independent
measurements (e.g. from neutrino factories or superbeams), the observation
of a large Earth effect in the antineutrino channel and of a smaller one
in the neutrino channel will testify for normal mass hierarchy and
differences between $\numu$ and $\nutau$ fluxes. Moreover, the character
of the $\numu$ -- $\nutau$ difference can be found from the sign of the
observed Earth effect: a positive (negative) effect at high energies will
correspond to $T_\tau > T_\mu$ or $T_\tau = T_\mu$ with $L_\tau > L_\mu$
($T_\tau < T_\mu$ or $T_\tau = T_\mu$ with $L_\tau < L_\mu$).   

If the same experimental result is observed but $P_H$ is unknown, an
ambiguity exists between the two possibilities: (1) $F^0_\mu\neq F^0_\tau$
with $P_H=0$, and (2) $F^0_\mu =  F^0_\tau $ with $0<P_H<1$. In some 
circumstances the ambiguity can be resolved by considering the sign of the
effect: if the effect is positive the case (1) will be established, with
$T_\tau > T_\mu$ or $T_\tau = T_\mu$ with $L_\tau > L_\mu$.  Moreover, the
information that $P_H\ll P^{eq}_H$ is obtained.
If the effect is negative, the case (1) is allowed only if $T_\tau <
T_\mu$ or $T_\tau = T_\mu$ with $L_\tau < L_\mu$.
Therefore a full discrimination requires the knowledge of the sign of
$\epsilon_T$ and $\epsilon_L$.

If a large Earth matter effect is observed in the $\nue$ spectrum together
with a small one in the $\barnue$ spectrum, the inverted hierarchy and the
existence of the $\barnumu$ -- $\barnutau$ flux differences will be
established, with implications analogous to those given here for the
neutrino channel.     
\\

\noindent
Let us summarize our main results.
\\

\noindent
1. The formalism of  flavor conversion of supernova neutrinos has been
generalized to the case of unequal fluxes of the non-electron neutrinos. 
A proof was given that, even for different fluxes of $\numu$ and
$\nutau$ neutrinos, the effects of CP violation are not observable 
in the supernova neutrino signal since they do not affect the observed
$\nue$  flux and  cancel in the sum of the $\numu$ and $\nutau$ fluxes in the 
detector.
\\

\noindent
2. It was found that possible $\numu$ -- $\nutau$ ($\barnumu$ -- $\barnutau$) 
flux differences modify the observed $\nue$ ($\barnue$) flux compared to  
what is predicted for equal fluxes if the mass hierarchy is normal (inverted) 
and the adiabaticity violation in the H resonance is not maximal, $P_H<1$. 
\\

\noindent
3. For conversion in the star only, the effect of these differences
consists in a distortion (broadening) of the observed $\nue$ or
$\barnue$ energy spectrum. 
For relative differences between the temperatures and/or integrated 
luminosities of the muon and tau neutrino fluxes $\lta 20\%$ the relative
deviation $R$ of the spectrum 
can be as large as $R\sim 0.4$ at high energies. However, the spectrum is 
consistent with the undistorted one  
within the  uncertainties due to 
the approximate 
%($\sim 10\%$) 
knowledge of the neutrino temperatures and 
luminosities.
\\

\noindent
4. If the solution of the solar neutrino problem is in the LMA region and
the neutrino trajectory crosses the Earth, oscillatory distortions of the
$\nue$ or $\barnue$ energy spectra occur. For $P_H\sim 1$ ( $\sin^2
\theta_{13}< 10^{-4}$) the effect of the $\numu$ -- $\nutau$ ($\barnumu$ 
-- $\barnutau$) flux differences is small and indistinguishable from the
equal fluxes case due to the uncertainties in the neutrino temperatures
and luminosities.
\\

\noindent
5. Conversely, if $P_H\simeq 0$ 
%%%($P_H\ll P^{eq}_H$) 
($\sin^2\theta_{13}\gta 10^{-3}$) and the hierarchy is
normal (inverted), the Earth matter effect in the neutrino (antineutrino) 
channel is entirely due to the $\numu$ -- $\nutau$ ($\barnumu$ -- $\barnutau$) 
flux difference and therefore the existence of the flux difference can, in
principle, be established. 
In the neutrino channel, the relative deviation $r$ of the observed
spectrum from the prediction in the case of no Earth crossing can be
typically as large as $r\simeq 0.2 - 0.3$ at high energy, $E\simeq 70-80$
MeV, and can reach $r\sim 0.7 $ for the largest considered values of the
difference $T_\tau-T_\mu$. This effect could be observable in future if large 
$\nue$ detectors with good energy resolution become available.  
In the antineutrino channel the effect does not exceed $\sim 10 - 15 \%$
and could be tested by the next generation of large Cherenkov detectors.    
\\

\noindent
6. The observation of the Earth matter effect that we considered here  
would be a signal of the difference between $\numu$ and  $\nutau$ (or
$\barnumu$ and $\barnutau$) fluxes, with interesting implications for the 
models of neutrino transport in the star and for the underlying physics.
%these models implement.
Jointly with other observations, it would give information on
the neutrino mass hierarchy and on the mixing angle $\theta_{13}$.
The effects of the possible $\numu$ -- $\nutau$ flux difference should be 
properly taken into account in the interpretations of the results of the
supernova neutrino experiments. 
\\

\noindent
7. The potential $V_{\tau\mu}$ is induced by radiative corrections and
therefore is much smaller than the potential $V_e$. Nevertheless, it
plays an important role in the evolution of the neutrino states in the
star provided that the original fluxes of $\numu$ and $\nutau$ are
different. Had one disregarded it, one would have arrived at a much
smaller Earth matter effect, suppressed by the factor $\cos
2\theta_{23}\ll 1$.

\subsection*{Acknowledgements} 
The authors wish to thank H.~T.~Janka and G.~G.~Raffelt for fruitful
discussions. C.L. thanks A.~Friedland, T.~Totani, J.~Klein, V.~Barger,
D.~Marfatia and B.~P.~Wood for useful comments and acknowledges
partial support of her work from the NSF grant PHY-0070928. E.A. was
supported by the Calouste Gulbenkian Foundation as a Gulbenkian
Visiting Professor at Instituto Superior T\'ecnico.

%%%%%%%%%%%%%  BIBLIOGRAPHY %%%%%%%%%%%%%%%%%%%%%%%%%%%%%%%%%%%%%%%%%%%%%%%

%\newpage

\bibliography{draft}

%%%%%%%%%%%%%%%%%%%%%%%%%%%%%%%%%%%%%%%%%%%%%%%%%%%%%%%%%%%%%%%%%%%%%%%%%%

%%%%%%%%%%%%%  FIGURES  %%%%%%%%%%%%%%%%%%%%%%%%%%%%%%%%%%%%%%%%%%%%%%%

\newpage

\end{document}